\documentclass[english,prd,aps,preprint,nofootinbib,superscriptaddress,showpacs]{revtex4}
\usepackage{longtable}
\usepackage{amsmath}
\usepackage{graphicx}
\usepackage{amssymb}
\usepackage{esint}
\usepackage{flafter}
\usepackage{dcolumn}
\usepackage{bm}
\usepackage{babel}
\def\bee{\begin{eqnarray}}
\def\eee{\end{eqnarray}}

\makeatletter

\providecommand{\tabularnewline}{\\}

\makeatother

\begin{document}

\title{$B$ to $\pi$ Form Factors in collinear factorization approach}

\author{Tsung-Wen Yeh}


\affiliation{Department of Science Application and Dissemination, National Taichung University,
140 Ming-Sheng Rd, 40305 Taichung, Taiwan, R.O.C.}

\email{twyeh@ms3.ntcu.edu.tw}

\begin{abstract}
The form factors for semi-leptonic $B$ decays, $\bar{B}\to\pi l\bar{\nu}_l$, are calculated 
under collinear factorization approach. 
The end-point divergences are regularized by a $\xi$-regularization, 
where $\xi$ means the collinear fraction of the spectator anti-quark of the $\bar{B}$ meson. 
The form factors are calculated up-to $O(\alpha_{s}/m_{B})$. 
The complete $O(1/m_{B})$ contributions from the $\pi$ meson are calculated explicitly by
a collinear expansion method. 
A well-defined power expansion scheme is built such that 
the $O(1/m_{B})$ contributions are about $30\%$ of the leading order contributions.
A small value $F_+(0)=0.164$ is found.  
This confirms the SCET result $F_+(0)=0.17$ from $B\to\pi\pi$ decays.
The form factors are calculated for {\normalsize $0\leq q^{2}\leq16\,\text{GeV}^{2}$},
where $q^2$ is the invariant mass of the lepton pair in $\bar{B}\to\pi l\bar{\nu}_l$.
An extrapolation of the form factors to $q^{2}>16\,\text{GeV}^{2}$ is made to obtain
{\normalsize $|V_{ub}|^{-2}\int_{0}^{26.42\,\text{GeV}^{2}}dq^{2}(d\Gamma/dq^{2})_{\text{th}}=(5.71\pm 0.91)ps^{-1}$}.
We determine $|V_{ub}|=(3.95\pm 0.13_{\text{exp}}\pm 0.32_{\text{th}})\times 10^{-3}$ 
from the world averaged branching ratio 
$Br(\bar{B}\to\pi l\bar{\nu}_{l})=(1.36\pm 0.09)\times 10^{-4}$.
\end{abstract}
\maketitle

\section{Introduction}
%
$B$ factories have obtained precise results on Cabibbo-Kobayashi-Maskawa
(CKM) matrix elements of the standard model (SM) \cite{DiLodovico:2008uw}. 
The LHCb has already started running. 
More results on CKM matrix elements with higher accuracy are expected in near future \cite{Eisenhardt:2008zz}. 
Being an important element of CKM matrix elements,
$|V_{ub}|$ still contains large uncertainties dominated by theoretical ones. 
For example, the $|V_{ub}|$ from inclusive $\bar{B}\to X_{u}l\bar{\nu}_{l}$ processes
contains $10\%$ uncertainties, where the $7\%$ uncertainty comes from the $60$ MeV uncertainty for $m_b$
and the other $3\%$ uncertainty comes from experiments. 
The $|V_{ub}|$ from exclusive $\bar{B}\to\pi l\nu_{l}$ processes, 
contains a $10-15\%$ uncertainty \cite{Khodjamirian:1997ub, Ball:1998tj,Khodjamirian:1998ji,Ball:2004ye,
Ball:2005tb,Khodjamirian:2006st,Duplancic:2008ix,Bowler:1999xn, Okamoto:2004xg,Shigemitsu:2004ft,Dalgic:2006dt,
Becirevic:1999kt,Fukunaga:2004zz,Arnesen:2005ez,Flynn:2007qd,Flynn:2007ii,Ball:2007sw} 
from the $B\to\pi$ form factor,
$F_{+}^{B\pi}$,
and a $6\%$ uncertainty from experiments
\cite{Athar:2003yg, Aubert:2005cd, Aubert:2005tm}.
In addition, the $q^2$ spectrum of $\bar{B}\to\pi l\nu_{l}$ has been well constrained experimentally 
\cite{Aubert:2005tm}.

To take full advantage of the experimental precision for exclusive $\bar{B}\to\pi l\nu_{l}$, 
it is necessary to pin down the theoretical uncertainties on $F_{+}^{B\pi}$ to few percentage level. 
However, it is still a difficult task. 
Currently, nonperturbative methods, 
including QCD light-cone sum rule (LCSR)
\cite{Khodjamirian:1997ub, Ball:1998tj,Khodjamirian:1998ji,Ball:2004ye,
 Ball:2005tb,Khodjamirian:2006st,Duplancic:2008ix} 
and lattice QCD (LQCD)
\cite{Bowler:1999xn, Okamoto:2004xg,Shigemitsu:2004ft,Dalgic:2006dt}, 
are available. 
Innovative parameterization (PA) methods with model independent inputs from theories
have been built 
\cite{Becirevic:1999kt,Fukunaga:2004zz,Arnesen:2005ez,Ball:2006jz,Flynn:2007qd,Flynn:2007ii}. 

Naive application of perturbative QCD (pQCD) to $F_{+}^{B\pi}$ needs to account for 
logarithmic or linear end-point divergences  
\cite{Isgur:1988iw,Szczepaniak:1990dt,Akhoury:1993uw, Khodjamirian:1998vk, Beneke:2000wa}.
Different methods have been proposed for the end-point divergences
\cite{Sterman:1986fn,Sterman:1986aj,Akhoury:1993uw,Beneke:2000wa, Kurimoto:2001zj, Li:2001ay, Wei:2002iu, 
Lu:2002ny, Huang:2004hw, Bauer:2004tj, Manohar2007}. 
In principle, the pQCD method can give a model independent determination for the $|V_{ub}|$ 
and its precision can be improved order by order by perturbation theories.
Based on factorization theorem \cite{Brodsky:1981rp}, 
the pQCD expression for the form factor can
be written into a factorized formula in terms of hard scattering function
and nonperturbative meson distribution amplitudes (DAs). 
The hard scattering function contains short distance contributions and can
be calculated, perturbatively. 
The meson DAs contain long distance contributions and are universal. 
Once the meson DAs are determined from other processes, 
the factorization formula can make model independent predictions. 
However, due to large uncertainties associated with the pQCD form factors
\cite{Kurimoto:2001zj, Li:2001ay,Wei:2002iu, Lu:2002ny, Huang:2004hw},  
the pQCD method has not been applied to derive equally precise $|V_{ub}|$ 
as the other methods, such as, LCSR, LQCD, and PA.
In this paper, we would like to improve the precision order of the pQCD form factors,
such that one can derive the $|V_{ub}|$ with equally theoretical uncertainties as the other approaches. 

There are two compatible pQCD methods, 
the collinear factorization denoted by $C^{f}$,
and the $k_{T}$ factorization, or , PQCD factorization, 
denoted by $k_{T}^{f}$.
The $k_{T}^{f}$ has been widely used for $F_{+}^{B\pi}$ 
\cite{Kurimoto:2001zj, Li:2001ay, Wei:2002iu, Lu:2002ny, Huang:2004hw}.
In $k_{T}^{f}$, end-point singularities are supposedly solved 
by the intervention of parton transverse momenta \cite{Li:1992nu}. 
However, the transverse parton momenta would induce large logarithms
$\alpha_{s}\ln^{2}k_{T}$ from higher loop corrections \cite{Kurimoto:2001zj}. 
In addition, there are also large logarithms $\alpha_{s}\ln^{2}x$ associated with subleading
twist (twist-3) contributions \cite{Kurimoto:2001zj}. 
These large logarithms need to be re-summed to be Sudakov factors \cite{Kurimoto:2001zj}.
The Sudakov factors are expected to suppress the contributions from end-point regions \cite{Wei:2002iu}. 
In practical applications, some criteria for the Sudakov factors are needed 
\cite{Li:1992nu, Kurimoto:2001zj,Wei:2002iu}. 
When $k_{T}^{f}$ is generalized to include subleading order contributions in the $1/m_{B}$ expansion, 
the subleading order, $O(1/m_{B})$, 
corrections dominate over the leading order contributions  \cite{Kurimoto:2001zj, Wei:2002iu}. 
Intrinsic transverse degrees of freedom of the
meson wave functions (for $B$ meson and pion ) are needed to cure
the ill behavior of the power expansion \cite{Huang:2004hw}. 

Unlike the complicate features and related issues of $k_{T}^{f}$, 
$C^{f}$ has a simple structure and is directly related to the parton model (PM)\cite{Lepage:1980fj}. 
It is expected that, 
once the end-point singularity can be regularized within $C^{f}$, 
the $C^{f}$ formalism would be instructive for both theory and phenomenology.
In the approach proposed by Akhoury, Sterman, and Yao (ASY) \cite{Akhoury:1993uw}, 
the heavy quark effective theory (HQET) and Sudakov re-summation were incorporated with $C^{f}$.
The end-point divergent problem is solved in the ASY approach.
However, a dynamical zero point was found and a small partial decay
rate for $\bar{B}\to\pi l\nu_{l}$ was obtained. 

In this paper, we propose a different approach  
to solve the end-point divergent problem 
and avoid the problems in the ASY approach. 
The key solution is a $\xi$-regularization (denoted by $\xi^{R}$) 
which can effectively regularize the end-point divergences. 
The $\xi^{R}$ has been applied to effectively regularize the end-point divergences in twist-3 hard spectator 
and annihilation contributions in charmless hadronic $B$ decays \cite{Yeh:2007fi}. 
In this paper, we show that $\xi^R$ is also effective for $F^{B\pi}_+$.
The suppression of end-point radiative corrections is provided by the $B$ meson distribution
amplitude (DA). 
This provides a stronger suppression effect than any Sudakov factors. 
The extension of application range of $q^{2}$ is given by including subleading corrections in $1/m_{B}$ expansion.
The complete $O(1/m_{B})$ contributions from the $\pi$ meson side
are calculated by using a collinear expansion (CE) method for exclusive processes,
which is developed by Yeh \cite{Yeh:2001ta, Yeh:2007fi, Yeh:2008xc}. 
The linear end-point divergences in the $O(1/m_{B})$ contributions are solved
by a simultaneous use of the $\xi^{R}$ and a non-constant twist-3 pseudo-scalar (PS) DA. 
The factorization of the $B\to\pi$ form factors has been shown valid under $C^f$ \cite{Kurimoto:2001zj}
and soft collinear effective theory (SCET)\cite{Bauer:2004tj}, respectively.
There lacked explicit regularization methods for the end point divergences in these previous proofs 
\cite{Kurimoto:2001zj, Bauer:2004tj}.
Our approach provides practical calculations for the form factors to show the factorization up to $O(\alpha_s/m_B)$.  
The non-constant twist-3 PS DA exists for a $\pi$ meson in its energetic state. 
On the other hand, 
a constant PS DA is usually used in the literature. 
As shown in \cite{Yeh:2007fi},
the constant PS DA is appropriate to describe a $\pi$ meson in its chiral state (, or, a soft pion),
but not an energetic pion. 
Unexpected large contributions associated with the constant PS DA are already noticed in $k_{T}^{f}$ 
\cite{Kurimoto:2001zj,Wei:2002iu}
and LCSR \cite{Belyaev:1993wp, Khodjamirian:1997ub, Ball:1998tj, Khodjamirian:1998ji,Duplancic:2008ix} .
If the constant PS DA corresponds to the soft pion state,
then these large contributions from the constant pion PS DA can be identified as soft dominated contributions 
instead of hard dominated contributions
as expected in the calculations performed in the $k_T^f$ and LCSR. 
This point of view of using the pion state corresponding to the DA to identify the kinds of contributions (soft or hard)
is different from the traditional way of using the scaling of the relevant contributions.
This provides another method to identify the considered contributions.
The unexpected large contributions from the pion PS DA in $k_T^f$ and LCSR could be overestimated.
Since end-point singularities at leading
and subleading order in $1/m_{B}$ expansion can be effectively solved, 
the $F_{+,0}^{B\pi}$ are calculable under $C^{f}$.  
Another effect associated with $\xi^R$ is that, for $\bar{u}\sim O(1)$, 
the divergences from $\eta\sim O(\Lambda/m_B)$ can be regularized.
This extends the application range of $C^f$ from small $q^2\sim 0$ 
to moderate $q^2\sim 16$ GeV$^2$. 
Note that the relevant energy scale of $\alpha_{s}$ is set as $t=1.65\sqrt{1-q^{2}/m_{B}^{2}}$,
which is about $1\,\text{GeV}$ at $q^{2}=16\,\text{GeV}^{2}$ and
the coupling constant $\alpha_{s}$ is about $0.48$, or, $\alpha_{s}/\pi\simeq0.16<1$.
There are also possible subleading order contributions in $1/m_{B}$
expansion from the $B$ meson side. 
Only $1/m_{B}$ contributions from the two parton Fock state of the $B$ meson  are calculated. 
In summary, we plan to make the following progresses in theory.
\begin{enumerate}
\item The $C^{f}$ is applicable for $F_{+,0}^{B\pi}$ at leading twist order by $\xi^{R}$.
\item $O(1/m_B)$ two parton contributions are shown calculable under $C^{f}$.
\item The twist-3 three parton corrections from the pion are explicitly calculated.
This is a first result in the literature. 
\item The complete twist-3 contributions
are shown less than the twist-2 contributions for $0\leq q^{2}\leq16\,\text{GeV}^{2}$. 
A well-defined power expansion is given.
\end{enumerate}

The result is applied to extract $|V_{ub}F^{B\pi}_+(0)|$ and $|V_{ub}|$ from the world averaged branching fraction 
$Br^{\text{exp}}_{\text{SL}}$ for semi-leptonic decays $\bar{B}\to\pi l\bar{\nu}_l$.
A fitting method is used to determine the value of a parameter $\omega_B$ for the $B$ meson distribution amplitude.
A parameterization form, $F^{C^f}_+(q^2)$, of $F^{B\pi}_+(q^2)$ versus $q^2$ is given by a minimal $\chi^2$ fitting.
Our analysis gives $|V_{ub}|=(3.95\pm 0.13_{\text{exp}}\pm 0.32_{\text{th}})\times 10^{-3}$ 
with experimental and theoretical errors.     
This agrees well with the world averaged value of $|V_{ub}|$, $|V_{ub}|=(3.95\pm 0.35)\times10^{-3}$ 
\cite{Amsler:2008zz}.
The fit form factor $F^{C^f}(q^2)$ predicts $F^{B\pi}_+(0)=0.16$
which confirms the founding $F^{\text{SCET}}_{+}(0)=0.17$ 
by the soft collinear effective theory (SCET) from an analysis for $B\to\pi\pi$ decays \cite{Bauer:2004tj}.

The organization is as follows. $\xi^{R}$ is defined and shown effective
for leading twist contributions in Section II. The comparison between
the $\xi^{R}$ and the $k_{T}$-regularization (denoted by $k_{T}^{R}$
in this paper) is given in this Section. To generalize $C^{f}$ for
higher twist contributions, the collinear expansion method is used
in Section III. In Section IV the $F_{+,0}^{B\pi}$ are explicitly
calculated up-to $O(\alpha_{s}/m_{b})$. In Section V, the form factors
are applied to determine $|V_{ub}|$ from experiments. Last two Sections
are devoted for comparisons and discussions. 
Two Appendices are given.

\section{Leading twist $B\to\pi$ form factors, end-point divergences, and
$\xi$-regularization}

In this section, we first review how the end-point divergent problem
of the leading twist $B\to\pi$ transition form factors can arise.
We then define the $\xi^{R}$ and explain how it is effective for
end-point singularity. The $B\to\pi$ form factors $F_{+,0}^{B\pi}(q^{2})$
are defined by 
\begin{equation}
\langle\pi(p_{\pi})|\bar{q}\gamma^{\mu}b|\bar{B}(P_{B})\rangle
= 2F_{+}^{B\pi}(q^{2})p_{\pi}^{\mu}+[F_{+}^{B\pi}(q^{2})-(F_{+}^{B\pi}(q^{2})-F_{0}^{B\pi}(q^{2}))\frac{(m_{B}^{2}-m_{\pi}^{2})}{q^{2}}]q^{\mu}\;,\label{eq:FF-001}
\end{equation}
where $q=P_{B}-p_{\pi}$. 
Another set of form factors, $f_{1,2}^{B\pi}$, is also used in literature. 
Their definitions are 
\begin{equation}
\langle\pi(p)|\bar{q}\gamma^{\mu}b|\bar{B}(P_{B})\rangle
=f_{1}^{B\pi}(q^{2})P_{B}^{\mu}+f_{2}^{B\pi}(q^{2})p_{\pi}^{\mu}\;.
\end{equation}
$F_{+,0}^{B\pi}$ and $f_{1,2}^{B\pi}$ are related by the following identities, 
\begin{eqnarray}
F_{+}^{B\pi}(q^{2}) & = & \frac{1}{2}(f_{1}^{B\pi}(q^{2})+f_{2}^{B\pi}(q^{2}))\,,\label{eq:f12toFFplus}\\
F_{0}^{B\pi}(q^{2}) & = & \frac{1}{2}((2-\eta)f_{1}^{B\pi}(q^{2})+\eta f_{2}^{B\pi}(q^{2}))\,,\label{eq:f12toFFzero}
\end{eqnarray}
where $\eta=1-q^{2}/m_{B}^{2}$. 
Under $q^{2}\to 0$, $F_{+}^{B\pi}(q^{2})=F_{0}^{B\pi}(q^{2})$.
At maximal recoil limit (the energetic limit for the $\pi$ meson), 
the form factors $F_{+}^{B\pi}(q^{2})$ and
$F_{0}^{B\pi}(q^{2})$ becomes identical to cancel the $q^{2}\to 0$
pole in Eq.(\ref{eq:FF-001}). 
The $B$ meson's momentum $P_{B}$ is defined in the $B$ meson's rest frame 
 $P_{B}^{\mu}=(m_{B},0,0,0)=(P_{B}^{+},P_{B}^{-},0_{\perp})$
with $P_{B}^{+}=P_{B}^{-}=m_{B}/\sqrt{2}$. 
The $\pi$ meson momentum $p_{\pi}$ is written as 
$p_{\pi}^{\mu}=\frac{1}{2}(\eta m_{B},0,0,\eta m_{B})=(p_{\pi}^{+},0,0_{\perp})$
where $p_{\pi}^{+}=\eta m_{B}/\sqrt{2}$. 
$q=P_{B}-p_{\pi}=(m_{B}(1-\eta/2),0,0,-\eta m_{B}/2)=(q^{+},q^{-},0_{\perp})$
where $q^{+}=(1-\eta)m_{B}/\sqrt{2}$ and $q^{-}=m_{B}/\sqrt{2}$.
The light-cone coordinate system will be used in this work. 
Under $C^f$, partons carry collinear fractions of external meson momenta. 
The spectator anti-quark of the $\bar{B}$ meson carries a momentum $l_{sp}^{\mu}=(0,l_{sp}^{-},0_{\perp})$
where $l_{sp}^{-}=\xi P_{B}^{-}=\xi m_{B}/\sqrt{2}$. The $b$ quark's
momentum is defined as $P_{b}^{\mu}=(P_{b}^{+},P_{b}^{-},0_{\perp})$
with $P_{b}^{-}=\bar{\xi}P_{B}^{-}=\bar{\xi}m_{B}/\sqrt{2}$ and $P_{b}^{+}=m_{b}^{2}/(\sqrt{2}\bar{\xi}m_{B})$.
The $b$ quark is defined on-shell. This is different from the usual
treatment in the literature that the $b$ quark is assumed almost
on-shell, $P_{b}^{2}\simeq m_{B}^{2}$. The quark $q$ inside the
$\pi$ meson is defined to carry a momentum $l_{q}=(l_{q}^{+},0^{-},0_{\perp})$
with $l_{q}^{+}=u\eta m_{B}/\sqrt{2}$. The anti-quark $\bar{q}$
inside the $\pi$ meson has a momentum $l_{\bar{q}}=(l_{\bar{q}}^{+},0^{-},0_{\perp})$
with $l_{\bar{q}}^{+}=\bar{u}\eta m_{B}/\sqrt{2}$. $\xi$ and $u$
are momentum fraction variables and $\bar{\xi}=1-\xi$ and $\bar{u}=1-u$.
$E=m_{B}/\sqrt{2}$ is used in the following text. 

\begin{figure}[t]
\includegraphics[scale=0.5]{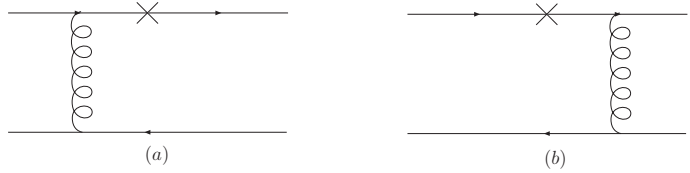}
\caption{One gluon exchanged Feynman diagrams for the hard scattering functions $H^{\mu}$ of the 
$C^f$ amplitude for the $\bar{B}\to \pi l\bar{\nu}_l$ decays. 
The external meson states $|\bar{B}\rangle$ and $\langle\pi|$ are not shown.
The cross vertex denotes the vector operator $\gamma^\mu$.
The diagram (a) describes the one gluon exchanged process 
between the bottom quark $b$ and the spectator anti-quark $\bar{q}$ of the $\bar{B}$ meson.
The diagram (b) describes the one gluon exchanged process 
between the quark $q$ and the anti-quark $\bar{q}$ of the $\pi$ meson.}
\label{fig:two parton}
\end{figure}

Under large recoil condition, $\eta\to1$, the $\pi$ meson has an
energetic momentum $p_{\pi}^{+}\simeq m_{B}/\sqrt{2}\gg m_{\pi}$.
We assume that the virtual gluon carries a hard-collinear energy scale. 
The PQCD is applicable because the involved strong coupling constant
$\alpha_{s}(t)$ at the interaction energy scale $t=\sqrt{\eta\Lambda_{h}m_{B}}$
with $\Lambda_{h}=0.5\,\text{GeV}$ is around $0.3$. 
Under $C^{f}$, the twist-2 contribution to the matrix element $\langle\pi|\bar{q}\gamma^{\mu}b|\bar{B}\rangle$
is written as 
\begin{eqnarray}
M^{tw2,\mu} & = & f_{\pi}f_{B}\frac{\pi\alpha_{s}(t)C_{F}}{N_{c}}
\int_{0}^{1}d\xi\phi_{B}^{tw2}(\xi)
\int_{0}^{1}du\phi_{\pi}^{P}(u)H^{tw2,\mu}(\xi,u)\,,
\end{eqnarray}
where $H^{tw2,\mu}(\xi,u)=H^{(a),tw2,\mu}(\xi,u)+H^{(b),tw2,\mu}(\xi,u)\,$
denote the hard scattering functions for the lowest order Feynman
diagrams depicted in Fig.~\ref{fig:two parton}(a) and (b). $\phi_{B}^{tw2}(\xi)$ and
$\phi_{\pi}^{P}(u)$ are the $B$ meson's and $\pi$ meson's leading
twist LCDA, respectively. $f_{B}$ and $f_{\pi}$ are the $B$ meson's
and $\pi$ meson's decay constants. $C_{F}=(N_{c}^{2}-1)/(2N_{c})$
and $N_{c}$ are color factors. $\alpha_{s}$ is the strong coupling
constant. $H^{(a,b),tw2,\mu}(\xi,u)$ are written as 
\begin{eqnarray}
H^{(a),tw2,\mu}(\xi,u) & = & 
\frac{1}{\eta\bar{u}\xi E^{2}}\left(P_{B}^{\mu}-\frac{1}{\eta}p_{\pi}^{\mu}\right)\,,
\end{eqnarray}
\begin{eqnarray}
H^{(b),tw2,\mu} & (\xi,u)= & 
\frac{(\bar{\xi}+\eta\bar{u})}{\eta\xi\bar{u}(\xi-\eta\bar{u})E^{2}}p_{\pi}^{\mu}\,.\label{eq:H(b)-tw2}
\end{eqnarray}
One can arrive at the form factors $f_{1,2}^{B\pi,tw2}$ 
\begin{eqnarray}
f_{1}^{B\pi,tw2}(q^{2}) & = & 
\frac{\pi\alpha_{s}(t)}{E^{2}}\frac{C_{F}}{N_{c}}f_{\pi}f_{B}
\int d\xi\phi_{B}^{tw2}(\xi)
\int du\phi_{\pi}^{P}(u)\left[\frac{1}{\eta\xi\bar{u}}\right]\,,\\
f_{2}^{B\pi,tw2}(q^{2}) & = & 
\frac{\pi\alpha_{s}(t)}{E^{2}}\frac{C_{F}}{N_{c}}f_{\pi}f_{B}
\int d\xi\phi_{B}^{tw2}(\xi)
 \int d u
  \phi_{\pi}^{P}(u)
   \left[
      \frac{(1+\eta)\eta\bar{u}-(1+\eta)\xi+\eta}{\eta^{2}\xi\bar{u}(\xi-\eta\bar{u})}
       \right]\,,
\end{eqnarray}
where $\bar{\eta}=1-\eta$ and $\bar{u}=1-u$. The form factors $F_{+,0}^{B\pi,tw2}$
from $f_{1,2}^{B\pi,tw2}$ by Eqs.(\ref{eq:f12toFFplus},\ref{eq:f12toFFzero})
are written as 
\begin{eqnarray}
F_{+}^{B\pi, tw2}(q^{2}) & = & \frac{\pi\alpha_{s}(t)}{2 E^{2}}
\frac{C_{F}}{N_{c}}f_{\pi}f_{B}
\int d\xi\phi_{B}^{tw2}(\xi)
\int d u\phi_{\pi}^{P}(u)
\left[\frac{\eta-\xi+\eta\bar{u}}{\eta^{2}\xi\bar{u}(\xi-\eta\bar{u})}\right]\,,
\end{eqnarray}
\begin{eqnarray}
F_{0}^{B\pi,tw2}(q^{2}) & = & 
\frac{\pi\alpha_{s}(t)}{2 E^{2}}
 \frac{C_{F}}{N_{c}}
  f_{\pi}f_{B}
  \int d\xi\phi_{B}^{tw2}(\xi)
  \int d u\phi_{\pi}^{P}(u)
    \left[
    \frac{\eta+(1-2\eta)\xi-(1-2\eta)\eta\bar{u}}{\eta\xi\bar{u}(\xi-\eta\bar{u})}
        \right]\,.
\end{eqnarray}

$\phi_{\pi}(u)$ is linear in $u$. 
$f_{1,2}^{B\pi,tw2}$ are finite.
However, if $(\xi-\eta\bar{u})$ is approximated to be $(-\eta\bar{u})$, 
then $f_{2}^{B\pi,tw2}$ becomes logarithmic divergence at $\bar{u}\to 0$.
This is the end-point divergent problem of the $B\to\pi$ transition form factors. 
The key point is that, without loss of generality, the
denominator of the internal $b$ quark propagator is approximately
to be, 
$(P_{b}+k)^{2}-m_{b}^{2}
\simeq m_{B}^{2}(\xi-\eta\bar{u})
+O(\Lambda_{\text{QCD}}\bar{\Lambda})
+O(\Lambda_{\text{QCD}}^{2})$
where $\bar{\Lambda}=m_{B}-m_{b}$, $\xi\simeq O(\Lambda/m_{B})$
and $\Lambda=\Lambda_{\text{QCD}}$. 
The error terms are of next-next-to-leading order in $1/m_{B}$ expansion. 
In the end-point region of $\bar{u}$, 
$\bar{u}\simeq O(\Lambda_{\text{QCD}}/m_{B})$, 
$\xi$ is as large as $\eta\bar{u}$ for $\eta\simeq O(1)$. 
$\xi$ retained in $(P_{b}+k)^{2}-m_{b}^{2}$ is necessary.

Akhoury, Sterman, and Yao (ASY) \cite{Akhoury:1993uw} proposed an approach to combine the heavy
quark effective theory (HQET), a resummation of Sudakov double logarithms,
and $C^{f}$ for exclusive processes.
In the ASY approach, the $b$ quark of the $B$ meson is almost static
at the energy scale much less than the $b$ quark's mass, $m_{b}$.
The exchanged gluon in the Feynman diagram in Fig.~\ref{fig:two parton} (b) carries soft
energy. The $b$ quark inside the $B$ meson can be effectively described
by an effective field, $h_{v}$, with $v=P_{B}/m_{B}$ of HQET. According
to HQET, the light degrees of freedom of the $B$ meson are identified
as a brown muck such that only soft interactions can exist between
the spectator anti-quark and the $b$ quark inside the $\bar{B}$
meson. There needs a subtraction operation to separate long distance
and short distance contributions from the $B$ meson side. The subtraction
operation is equivalent to neglect $\xi$ in the denominator and $\bar{\xi}$
in the numerator of the $H^{(b),tw2,\mu}$ in Eq.(\ref{eq:H(b)-tw2}).
It is obvious that once $\xi$ and $\bar{\xi}$ are subtracted from
the $H^{(b),tw2,\mu}$ term, the end-point singularity in $f_{2}^{B\pi,tw2}$
is solved. When loop corrections are concerned, there would arise
large infrared (IR) logarithms from the internal $b$ quark line in
the Feynman diagram in Fig. 1(b). A resummation over the large IR
logarithms gives Sudakov form factors. Although the ASY approach can
successfully solve the end-point divergent problem, there exist two
problems. The first one is that there is a dynamical zero point at
$\eta=1/2$ in the $|f_{1}+f_{2}|$. We find that the dynamical zero
point is due to the neglect of the $\bar{\xi}$ factor in the $H^{(b),tw2,\mu}$
term. Refer to Eq.(\ref{eq:H(b)-tw2}). The $\bar{\xi}$ factor is
recovered as one include the dynamical contributions from the $b$
quark. This can be understood by referring to Appendix \ref{sec:Reduced-hard-scattering}.
The dynamical zero point is avoided in $\xi^{R}$. The second is that
small partial rates for $B\to\pi l\bar{\nu_{l}}$ were predicted in
the ASY approach. 
This implies that the ignored contributions in the ASY approach are important. 
In $\xi^{R}$, the ignored contributions in the ASY approach are recovered.
See Section V for this point. 
In our approach, two problems in the ASY approach are solved.

In $k^f_T$, the divergent term $(\eta\bar{u})^{-1}$ is replaced by $(k_{T}^{2}+\eta\bar{u}E^{2})^{-1}$.
While the transverse momenta of partons are remained in the parton propagators, 
the partons become off-shell because only collinear and transverse momenta are kept. 
Off-shell partons may radiate infinite soft gluons as they pass through space 
before they compose into external mesons. 
Re-summation of soft gluon radiations results in Sudakov factors,
which are expected to suppress the soft radiations from the end-point region. 
However, some criteria are required. 
This is an uncertainty of $k_{T}^{f}$. 

$\xi$ is of order $O(\Lambda/E)$ that the $B$ meson distribution
amplitude $\phi_{B}(\xi)$ has a peak at $\xi\sim O(\Lambda/E)$.
In the end-point region of $\bar{u}$, $\bar{u}\sim O(\Lambda/E)$,
$\phi_{\pi}^{P}(u)\sim O(\Lambda/E)$. 
The overall scale of the leading
part of the form factor $f_{2}^{B\pi,tw2}$ is counted as $O(1/E^{2})$
times the decay constants $f_{\pi}f_{B}$. 
Similarly, $f_{1}^{B\pi,tw2}$ is dominated by large $\bar{u}\sim O(1)$ and is counted as $O(1/E^{2})$
times the decay constants $f_{\pi}f_{B}$. 
Although $f_{1}^{B\pi,tw2}$ and $f_{2}^{B\pi,tw2}$ receive contributions from different configuration
regions, they are of the same order. 
Accordingly, contributions of subleading order in $1/m_{B}$ expansion should be included.
They are calculated in next Section. 

The $\pi$ meson's twist-2 DA $\phi_{\pi}^{P}(u)$ is modeled to be
its asymptotic form $\phi_{\pi}^{P}(u)=6u\bar{u}$ by neglecting its
scale dependence. 
This is fulfilled for the precision of $O(\alpha_s)$. 
The $B$ meson's twist-2 DA $\phi_{B}^{tw2}(\xi)$
is assumed to be modeled as \cite{Grozin:1996pq}
\begin{eqnarray}
\phi_{B}^{tw2(I)}(\xi) 
& = & 
  \frac{m_{B}^{2}\xi}{\omega_{B}^{2}}
    \exp(-\frac{m_{B}}{\omega_{B}}\xi)
\end{eqnarray}
which satisfies the following conditions 
\begin{eqnarray}
\int_{0}^{\infty}d\xi
  \phi_{B}^{tw2(I)}(\xi) 
& = & 1
\,,\label{eq:nom-B-LCDA}\\
\int_{0}^{\infty}d\xi
  \frac{\phi_{B}^{tw2(I)}(\xi)}{\xi} 
& = & 
  \frac{m_{B}}{\lambda_{B}}
\,.\nonumber 
\end{eqnarray}
The value of $\lambda_{B}$ satisfies the condition 
$6\lambda_{B}\leq4\bar{\Lambda}$
with $\bar{\Lambda}=m_{B}-m_{b}$. 
Different models for the 
$\phi_{B}^{tw2}(\xi)$
have only tiny differences if they are constrained by Eq.(\ref{eq:nom-B-LCDA})
\cite{Lee2005}. We note that the integration range over $\xi$ in
the calculation of the form factors is $[0,1]$ instead of $[0,\infty]$.
The difference between these two integration ranges is equal to 
\begin{eqnarray*}
\vartriangle 
& = & 
\int_{0}^{\infty}d\xi
  \phi_{B}^{tw2,I}(\xi)
  - \int_{0}^{1}d\xi
     \phi_{B}^{tw2,I}(\xi)
\\
& = & \left(
       \frac{m_{B}}{\omega_{B}}
        +
         1
       \right)
       \exp[-\frac{m_{B}}{\omega_{B}}]
\nonumber \\
&& \simeq 1.29\times 10^{(-4)}
\,, \nonumber
\end{eqnarray*}
where the last estimated number is calculated by using $\omega_{B}=0.46\,\text{GeV}$
and $m_{B}=5.28\,\text{GeV}$. 
$\vartriangle$ is negligible. 

To distinguish from the $k_{T}^{R}$ for end-point divergences, we
name the retain of $\xi$ in the internal $b$ quark propagator as
the $\xi$-regularization, $\xi^{R}$. 
It is instructive to see how
$\xi^{R}$ works for end-point divergences. 
Let's consider the $\eta\to1$
limit of 
$F_{+}^{B\pi,tw2}(q^{2})$. 
Except of the relevant parameters,
the most singular part is the integrations 
\[
\int_{0}^{1}d\xi
\phi_{B}^{tw2}(\xi)
\int_{0}^{1}d u
\phi_{\pi}^{P}(u)
\frac{1}{\xi\bar{u}(\xi-\bar{u})}
\,.
\]
It is seen that once $\xi$ is neglected in $(\xi-\bar{u})^{-1}$,
an end-point divergence arise to be $\log(\bar{u})$. The retain of
$\xi$ results in, for the integration over $u$, 
\begin{eqnarray*}
\int_{0}^{1}d u
  \phi_{\pi}^{P}(u)
   \frac{1}{\bar{u}(\xi-\bar{u})} 
& = & 6 
       ( 
        1 
         + 
          \bar{\xi}
           \ln\frac{\xi}{\bar{\xi}}
          - i\pi\bar{\xi}
        )
\,.
\end{eqnarray*}
One can observe that the original logarithmic divergence $\log\bar{u}$
is indeed regularized by $\xi$, effectively. The further integration
over $\xi$ is analytic because the end-points for 
$\xi\to0$ 
or 
$\bar{\xi}\to 0$
are prevented by the $B$ meson distribution amplitude 
$\phi_{B}^{tw2}(\xi)$.
For example, the integration for the most singular term is analytical
as 
\begin{eqnarray*}
\int_{0}^{1}d\xi
\phi_{B}^{tw2(I)}(\xi)
\frac{\bar{\xi}}{\xi}
\ln \frac{\xi}{\bar{\xi}} 
& = & -
       \left[
        1-e^{-a}+(a-1)(\gamma_{E}+\Gamma(a)+\ln a)
        \right]
\,,
\end{eqnarray*}
where 
$a=m_{B}/\omega_{B}$, 
$\gamma_{E}$ 
is the Euler number, 
and
$\Gamma(a)$ 
is the Gamma function. 
Another widely used model for
$\phi_{B}^{tw2}(\xi)$ 
has the form 
\begin{eqnarray*}
\phi_{B}^{tw2(II)}(\xi) 
& = & 
    \frac{N_{B}\xi^{2}(1-\xi)^{2}}{(\xi^{2}+\epsilon_{B}(1-\xi)^{2})^{2}}
\,,
\end{eqnarray*}
where $N_{B}$ and $\epsilon_{B}$ are determined by Eq.(\ref{eq:nom-B-LCDA}).
Although the integration can not be expressed explicitly, the result
is also analytic because the $\phi_{B}^{tw2,II}$ has a stronger suppression
effect on the end-points, $\xi$, $\bar{\xi}\to0$. Of course, the
final result would depend on the model for $\phi_{B}^{tw2}(\xi)$.
This model dependence is due to our rare knowledge for the $B$ meson.
However, these two models give almost the same numerical results for $F^{B\pi,tw2}_+$ in $\xi^R$.
For example, we can use $\omega_B=0.46$ GeV to obtain
\begin{eqnarray}
F^{B\pi,tw2,(I)}_+(0)=0.110
\,,\nonumber\\
F^{B\pi,tw2,(II)}_+(0)=0.104
\,,
\end{eqnarray}
where $F^{B\pi,tw2,(I)}_+(0)$ and 
$F^{B\pi,tw2,(II)}_+(0)$ are calculated 
by using 
$\phi^{(I)}_B(\xi)$ and 
$\phi^{(II)}_B(\xi)$, respectively.
$N_{B}=0.1536$ and 
$\epsilon_{B}=0.0061$ 
are used in the calculation for $F^{B\pi,tw2,(II)}_+(0)$.

This shows explicitly that the $\xi^{R}$ indeed regularizes the end-point
divergence $\log\bar{u}$. 
The suppression of contributions from the
end-point region in $\xi^{R}$ is provided by the $B$ meson's DA,
$\phi_{B}^{tw2}(\xi)$. 
The success of $\xi^{R}$ implies that the dynamical role of the spectator
(anti-)quark of the $\bar{B}$ meson is important in the studies of
heavy to light processes. 
This is contrary to the heavy to heavy processes,
such as $\bar{B}\to Dl\nu$, in which heavy quark symmetry is useful.
In the heavy quark infinite limit of the heavy to heavy processes,
the light degrees of freedom of the heavy mesons remained independent
of the $h_{v}\to h_{v^{\prime}}l\nu$ process, 
described by a Isgure-Wise function, 
$\zeta(v\cdot v^{\prime})$. 
$h_{v(v^{\prime})}$ are effective fields 
for the $b$ and $c$ quarks of the $B$ meson and the $D$ meson, respectively. 
If the same picture applies for $\bar{B}\to\pi l\nu$ processes, 
then the $b\to ul\nu$ transition should be in a similar
condition as that of $h_{v}\to h_{v^{\prime}}l\nu$ process. 
This means that the final state $\pi$ meson should be in its soft pion state, 
at which the pQCD method is inapplicable in principle. 
In fact,
the light degrees of freedom of the $\bar{B}$ meson would experience
violate fluctuations during the $b\to ul\nu$ transition proceeds.
This implies the applicability of pQCD \cite{Hill:2005ju}. 
The heavy quark symmetry is inapplicable in $\bar{B}\to\pi l\bar{\nu}_l$ decays.
We will see later that the intervention of the $\xi$ variable becomes
a straightforward step under CE. See the next Section.

\section{Collinear expansion}

\begin{figure}[t]
\includegraphics[scale=0.5]{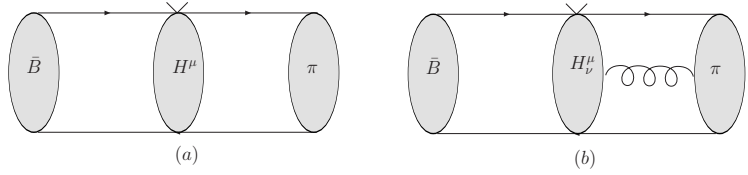}
\caption{Feynman diagrams for the hard scattering functions $H^{\mu}$ and $H^{\mu}_{\nu}$ of the 
$C^f$ amplitudes for the $\bar{B}\to \pi l\bar{\nu}_l$ decays. 
The external meson states $|\bar{B}\rangle$ and $\langle\pi|$ are shown by bubbles with symbols $\bar{B}$ and $\pi$, respectively.
The central bubble represents parton virtual interactions between relevant partons from
the $\bar{B}$ meson and pion. 
The cross vertex denotes the vector operator $\gamma^\mu$.
The diagram (a) describes the interaction processes with four partons involved.
The diagram (b) describes the interaction processes with five partons involved. 
There are the two quark partons and one gluon parton of the pion taking participate in the interactions.}
\label{fig:fig2}
\end{figure}

In this section, we describe how CE can be applied
to derive twist-3 two parton and twist-3 three parton contributions.
In this work, we consider only the twist-3 three parton contributions
from the $\pi$ meson. The three parton contributions from the $B$
meson are complicate and left to other places. 
We begin with the amplitude
for the diagrams in Fig.~\ref{fig:fig2},
\begin{eqnarray}
M^{\mu} 
& = & 
\int\frac{d^{4}l_{B}}{(2\pi)^{4}}
\int\frac{d^{4}l}{(2\pi)^{4}}
\text{Tr}
  [ 
    H^{\mu}(l_{B},l)
    \Phi_{B}(l_{B})
     \Phi_{\pi}(l)
    ]
\label{eq:amplitudefull}\\
&  & + 
      \int\frac{d^{4}l_{B}}{(2\pi)^{4}}
      \int\frac{d^{4}l}{(2\pi)^{4}}
      \int\frac{d^{4}l^{\prime}}{(2\pi)^{4}}
       \text{Tr}
        [
         H_{\nu}^{\mu}(l_{B},l,l^{\prime})
          \Phi_{B}(l_{B})
          \Phi_{\pi}^{\nu}(l,l^{\prime})
         ]
\nonumber \\
 &  & +  \cdots
\,,\nonumber 
\end{eqnarray}
where the first line corresponds to the four parton interactions (for
the diagram depicted in Fig.~\ref{fig:fig2}(a)), 
and the second line represent five
parton interactions (for the diagrams depicted in Fig.~\ref{fig:fig2}(b)). 
The higher Fock state interactions are neglected. 
The 
$H^{\mu}(l_{B},\, l)$,
and 
$H_{\nu}^{\mu}(l_{B},\, l,\, l^{\prime})$ 
describe the parton
interactions in the hard scattering center. 
The 
$\Phi_{B}(l_{B})$,
$\Phi_{\pi}(l)$, 
$\Phi_{\pi}^{\nu}(l,l^{\prime})$, 
are the hadron functions defined as 
\begin{eqnarray}
\Phi_{B}(l_{B}) 
& = & 
   \int d^{4}z 
     e^{il_{B}\cdot z}
    \langle 0|
      \bar{q}(z)b(0)
        |\bar{B}\rangle
\,,\\
\Phi_{\pi}(l) 
& = & 
    \int d^{4}z e^{i l\cdot z}
     \langle\pi|
       \bar{q}(z)q(0)
       |0\rangle
\,,\\
\Phi_{\pi}^{\nu}(l,l^{\prime}) 
& = & 
    \int d^{4}z 
     e^{i l\cdot z}
     e^{i(l^{\prime}-l)\cdot z^{\prime}}
     \langle\pi|
       \bar{q}(z)(-g)A^{\nu}(z^{\prime})q(0)
       |0\rangle
\,,
\end{eqnarray}
where color and spin indices are not shown explicitly. 
The gauge links between particle fields in the matrix elements are implicitly understood.
The trace operation $\text{Tr}$ is taken over the spin and color indices.

The factorization of the amplitude into partonic and hadronic parts is composed of three steps, 
the factorization of loop parton momenta, 
the color index factorization, 
and spin index factorization. 
The factorization of loop parton momenta is performed by means of a Taylor expansion for the partonic part
and followed by relevant integral transformations. 
The color index factorization and the spin index factorization are similar and can be done by color
algebra and Fierz identities, respectively. 
These three steps are shown explicitly below.

Once the partonic part is separated from the hadronic part of the amplitude,
it is an important task to examine whether the partonic part suffers from soft divergences.
This is the proof of the factorization theorem. 
The $O(\alpha_{s})$ analysis for the twist-2
and twist-3 two parton contributions for $B\to\pi$ form factors has been given in \cite{Nagashima:2002iw}. 
The factorizability of $B\to\pi$ form factors
under $C^{f}$ is shown valid upto twist-3 order under two parton approximation. 
An all order proof of the factorization theorem for $B\to\pi$ form factors is shown valid at twist-2 order in SCET \cite{Bauer2003}.
The factorizability of the twist-3 three parton contributions
to the $B\to\pi$ form factors is still inaccessible. 
We assume that this is also valid in this work. 
A simple analysis shows that the twist-3 three parton contributions are also factorizable in $C^{f}$.
However, the complete analysis is tedious and very technical. 
We skip this part and leave it for other place. 
In next section, the twist-3 three parton contributions from the pion are calculated based on this assumption. 
The analysis of the factorizability for the subleading
twist contributions from the $B$ meson is more complicate. 
The complete $O(\alpha_{s})$ factorization analysis also needs a lot of space
and is not appropriate to be given here. 
Similarly, we also assume that the factorization is also valid for the subleading twist contributions from the $B$ meson. 
The calculations of the relevant quantities are based on this assumption. 
At order $O(\alpha_{s})$, the explicit results given in the next section for the considered contributions
confirm the factorizability. 
However, a complete analysis for the subleading twist contributions is important to make sure that the
calculations given in this paper are perturbatively meaningful. 

In Eq.(\ref{eq:amplitudefull}), the loop parton momenta of the partons
of the $B$ meson, $\, l_{B}$, is defined to flow from the $B$ mesons
into the scattering center, and those momenta $l$ and $l^{\prime}$
are defined to flow from the scattering center into the $\pi$ meson.
To perform collinear expansion, the momenta, $l,\, l^{\prime},\, l_{B}$
are found convenient to separate their on-shell part from their off-shell
parts. 
For example, $l$ can be written as 
\begin{eqnarray}
l^{\mu} & = & 
\hat{l}^{\mu}
+\frac{l^{2}+l_{\perp}^{2}}{2n\cdot\hat{l}}n^{\mu}
+l_{\perp\,,}^{\mu}\label{eq:collinear momentum}
\end{eqnarray}
where $l^{2}$ denotes the virtuality of $l$, $l_{\perp}^{\mu}$
is the transverse momentum, and $l_{\perp}^{\mu}l_{\perp,\mu}=-l_{\perp}^{2}$.
$\hat{l}^{2}=n^{2}=0$. 
$n$ is an auxiliary light-like vector. 
Under $C^{f}$, only collinear partons can involve in hard scattering center.
The most important contributions should come from collinear momentum
configuration according to the power counting rules \cite{Yeh:2008xc,Yeh:2007fi}. 
If $l$ represents the momentum of a collinear parton,
i.e., $l$ is a collinear momentum, 
then $l$ has a limited (transverse) virtuality, 
$l^{2}\sim l_{\perp}^{2}\sim O(\Lambda^{2})$. 
In $C^{f}$, the partons external to the hard scattering center are on-shell. 
According to Eq.(\ref{eq:collinear momentum}), 
the on-shell parton can have a momentum $\hat{l}^{\mu}$ with $\hat{l}^{2}=0$, or, 
\begin{eqnarray}
l_{L}^{\mu} & = & \hat{l}^{\mu}+\frac{l_{\perp}^{2}}{2n\cdot\hat{l}}n^{\mu}+l_{\perp}^{\mu}\,,\\
l_{L}^{2} & = & 0\,.\nonumber 
\end{eqnarray}
The $\hat{l}^{\mu}$ is the collinear component of $l^{\mu}$ and
is used for collinear partons in $C^{f}$. 
The $l_{L}^{\mu}$ has non-collinear momenta and is usually used in the $k_{T}^{f}$.
It is noted that $n\cdot\hat{l}=uE$ with $u$ the momentum fraction
and $E$ the energy scale. 
For $u\sim O(1)$, 
the $n^{\mu}$ 
term of 
$l_{L}^{\mu}$ 
is suppressed than the 
$\hat{l}^{\mu}$ 
and 
$l_{\perp}^{\mu}$ 
terms. 
For 
$u\sim O(\Lambda/E)$, 
three parts of 
$l_{L}^{\mu}$ 
are equally important. 
There requires some cares for the use of 
$l_{L}^{\mu}$ 
in the 
$k_{T}^{f}$. 
Especially, it is important in the end-point region. 
We argue that the 
$n^{\mu}$ 
term of 
$l_{L}^{\mu}$ 
should be kept in 
$k_{T}^{f}$. 
The calculations for 
$B\to\pi$ 
form factors in 
$k_{T}^{f}$ 
should be rechecked.

The CE is able to derive higher twist contributions from higher Fock state, 
non-collinear components of parton momentum, 
and miss-matched
spin states. 
The higher twist contributions can be factorized into
partonic and hadronic parts, 
which are separately gauge invariant.
The parton model interpretation for higher twist contributions is
similar to the leading twist contributions.

The CE may confront with loop expansion (LE). We choose our strategy
\cite{Yeh:2008xc,Yeh:2007fi} to firstly perform the LE for the amplitudes
and then to use CE. 
Changing the application order of these two expansions
would not make differences. 
The parton functions are expanded in LE as 
\begin{eqnarray}
H^{\mu}(l_{B},l) 
& = & 
  H^{(1),\mu}(l_{B},l)
 + O(\alpha_{s}^{2})
\,,\\
H_{\nu}^{\mu}(l_{B},l,l^{\prime}) 
& = & 
   H_{\nu}^{(1),\mu}(l_{B},l,l^{\prime})
 + O(\alpha_{s}^{2})
\,,
\end{eqnarray}
where 
$H^{(1),\mu}$ 
and 
$H_{\nu}^{(1),\mu}$ 
are of order 
$O(\alpha_{s})$.
In this paper, 
we only consider 
$O(\alpha_{s})$ 
contributions and assume 
$O(\alpha_{s}^{2})$ 
not important.

The first step is to expand 
$H^{(1),\mu}(l_{B},l)$ 
and 
$H_{\nu}^{(1),\mu}(l_{B},l,l^{\prime})$
with respect to the collinear momenta, 
$\hat{l}_{B}$, $\hat{l}$,
and 
$\hat{l}^{\prime}$, 
by a Taylor expansion
\begin{eqnarray}
H^{(1),\mu}(l_{B},l) 
& = & 
 H^{(1),\mu}(\hat{l}_{B},\hat{l})
 + \sum_{k=l,l_{B}}
     \frac{\partial H^{(1),\mu}}{\partial l^{\nu}}|_{\left\{ k=\hat{k},k=l_{B},\, l\right\} }
      (l-\hat{l})^{\nu}
  + \cdots
  \,,\\
H_{\nu}^{(1),\mu}(l_{B},l,l^{\prime}) 
& = & 
   H_{\nu}^{(1),\mu}(\hat{l}_{B},\hat{l},\hat{l}^{\prime})
  + \sum_{k=l,l^{\prime}}
      \frac{\partial H_{\nu}^{(1),\mu}}{\partial k^{\eta}}|_{\left\{ k=\hat{k},k=l_{B},\, l,\, l^{\prime}\right\} }
      (k-\hat{k})^{\eta}
  + \cdots
   \,,
\end{eqnarray}
The choice of $\hat{l}_{B}=\bar{n}\cdot l_{B}$ is according to $\hat{l}^{(\prime)}=n\cdot l^{(\prime)}$
to make the internal virtual gluon carry most violate energy. The
Taylor expansion follows from the argument of the $C^{f}$ that the
hard scattering functions contain only collinear momenta of partons
\cite{Yeh:2007fi}. The co-variant gauge, $\partial\cdot A=0,$ is used.
Up-to twist-3 order, one only needs to consider the first term of
the above first expansion series and the first two terms in the above
second expansion series. 
By substituting them into the original convolution integrations, 
one can obtain 
\begin{eqnarray}\label{eq:leading-amplitude}
M^{\mu} & = & 
\int d\xi
 \int du
  \text{Tr}[H^{(1),\mu}(\xi,u)\phi_{B}(\xi)\phi_{\pi}(u)]\\
 &  & 
+\int d\xi
  \int du
   \int du^{\prime}
    \text{Tr}[H_{\nu}^{(1),\mu}(\xi,u,u^{\prime})\phi_{B}(\xi)\phi_{\pi}^{\nu}(u,u^{\prime})]\nonumber \\
 &  & 
+\int d\xi
  \int du
   \int du^{\prime}
    \text{Tr}[H_{\nu\eta}^{(1),\mu}(\xi,u,u^{\prime})\phi_{B}(\xi)
             \omega_{\eta^{\prime}}^{\eta}\phi_{\pi}^{\eta^{\prime}\nu}(u,u^{\prime})]\nonumber \\
 &  & +\cdots
\,,\nonumber 
\end{eqnarray}
where 
$\omega_{\eta^{\prime}}^{\eta}
= g_{\eta^{\prime}}^{\eta}
 -n_{\eta^{\prime}}\bar{n}^{\eta}$.
Two light-like auxiliary vectors are used 
$\bar{n}^{\mu}=(\bar{n}^{+},\bar{n}^{-},\bar{n}_{\perp})=(1,0,0)$,
$n^{\mu}=(n^{+},n^{-},\bar{n}_{\perp})=(0,1,0)$. 
The hard scattering functions 
$H^{(1),\mu}(\xi,u)$, 
$H_{\nu}^{(1),\mu}(\xi,u,u^{\prime})$,
$H_{\nu\eta}^{(1),\mu}(\xi,u,u^{\prime})$ 
are defined by low energy theorems 
\begin{eqnarray}
H^{(1),\mu}(\xi,u) & = & H^{(1),\mu}(\hat{l}_{B},\hat{l})\,,\\
H_{\nu}^{(1),\mu}(\xi,u,u^{\prime}) & = & H_{\nu}^{(1)\mu}(\hat{l}_{B},\hat{l},\hat{l}^{\prime})\,,\\
H_{\nu\eta}^{(1),\mu}(\xi,u,u^{\prime}) & = & 
(\frac{\partial H_{\nu}^{(1),\mu}}{\partial l^{^{\prime}\eta}}-
\frac{\partial H_{\nu}^{(1),\mu}}{\partial l^{\eta}})|_{(l_{B}=\hat{l}_{B},l=\hat{l},l^{\prime}=\hat{l}^{\prime})}\,.
\end{eqnarray}
The meson functions are defined as 
\begin{eqnarray}
\phi_{B}(\xi) & = & 
\int\frac{d^{4}l_{B}}{(2\pi)^{4}}
 \int d^{4}z
   \delta(\xi-\frac{l_{B}^{-}}{P_{B}^{-}})
    e^{i l_{B}\cdot z}\langle0|\bar{q}(z)b(0)|B\rangle\,,\\
\phi_{\pi}(u) & = & 
  \int\frac{d^{4}l}{(2\pi)^{4}}
   \int d^{4}z
    \delta(u-\frac{l^{+}}{p_{\pi}^{+}})
     e^{i l\cdot z}\langle\pi|\bar{q}(z)q(0)|0\rangle\,,\\
\phi_{\pi}^{\nu}(u,u^{\prime}) & = & 
   \int\frac{d^{4}l}{(2\pi)^{4}}
    \int\frac{d^{4}l^{\prime}}{(2\pi)^{4}}
     \int d^{4}z\int d^{4}z^{\prime}
      e^{i l\cdot z}e^{i(l^{\prime}-l)\cdot z^{\prime}}\\
 &  & \times \delta(u-\frac{l^{+}}{p_{\pi}^{+}})
      \delta(u^{\prime}-\frac{l^{\prime+}}{p_{\pi}^{+}})
       \langle\pi|\bar{q}(z)(-g)A^{\nu}(z^{\prime})q(0)|0\rangle\,,\\
\phi_{\pi}^{\eta\nu}(u,u^{\prime}) & = & 
   \int\frac{d^{4}l}{(2\pi)^{4}}
    \int\frac{d^{4}l^{\prime}}{(2\pi)^{4}}
     \int d^{4}z\int d^{4}z^{\prime}
      e^{i l\cdot z}e^{i(l^{\prime}-l)\cdot z^{\prime}}\\
 &  & \times \delta(u-\frac{l^{+}}{p_{\pi}^{+}})
              \delta(u^{\prime}-\frac{l^{\prime+}}{p_{\pi}^{+}})
               \langle\pi|\bar{q}(z)(-i g)G^{\eta\nu}(z^{\prime})q(0)|0\rangle\,.
\end{eqnarray}
By using the identities 
\begin{eqnarray}
\delta(u-\frac{l^{+}}{p_{\pi}^{+}}) & = & 
  \int\frac{d\lambda}{(2\pi)}e^{i\lambda(u-\frac{n\cdot l}{E})}\,,\\
\int\frac{d^{4}l}{(2\pi)^{4}}\int d^{4}z 
  e^{i l\cdot(z-\frac{\lambda}{E}n)} & = & 
  \delta^{(4)}(z-\frac{\lambda}{E}n)\,,
\end{eqnarray}
and their similarities for the corresponding integrations, 
the meson functions become
\begin{eqnarray}
\phi_{B}(\xi) & = & 
  \int\frac{d\lambda}{(2\pi)}
   e^{i\lambda\xi}
    \langle0|\bar{q}(\frac{\lambda}{E}\bar{n})b(0)|B\rangle\,,\\
\phi_{\pi}(u) & = & 
  \int\frac{d\lambda}{(2\pi)}
    e^{i\lambda u}
     \langle\pi|\bar{q}(\frac{\lambda}{E}n)q(0)|0\rangle\,,\\
\phi_{\pi}^{\nu}(u,u^{\prime}) & = & 
  \int\frac{d\lambda}{(2\pi)}
   \int\frac{d\lambda^{\prime}}{(2\pi)}
     e^{i\lambda u} e^{i\lambda^{\prime}(u^{\prime}-u)}
    \langle\pi|
        \bar{q}(\frac{\lambda}{E}n)
          (-g)A^{\nu}(\frac{\lambda^{\prime}}{E}n)
            q(0)
             |0\rangle
     \,,\\
\phi_{\pi}^{\eta\nu}(u,u^{\prime}) & = & 
   \int\frac{d\lambda}{(2\pi)}
    \int\frac{d\lambda^{\prime}}{(2\pi)}
      e^{i\lambda u} e^{i\lambda^{\prime}(u^{\prime}-u)}
     \langle\pi|\bar{q}(\frac{\lambda}{E}n)(-i g)G^{\eta\nu}(\frac{\lambda^{\prime}}{E}n)q(0)|0\rangle\,.
\end{eqnarray}
where $G^{\eta\nu}=\partial^{\eta}A^{\nu}-\partial^{\nu}A^{\eta}$.
We note that the coupling $g$ is absorbed into the meson functions, 
$\phi_{\pi}^{\nu}$ and $\phi_{\pi}^{\eta\nu}$. 
The three parton contributions are then
counted as the same order of the two parton ones. 
This is different from the counting rule in $k_{T}^{f}$ \cite{Kurimoto:2001zj}, 
in which the three parton contributions are counted as
one more $O(\alpha_{s})$ order than the two parton ones. 

The second term in Eq.(\ref{eq:leading-amplitude}) is related to gauge phase factors
\begin{eqnarray}
 && \int d\xi
   \int d u
    \int d u^{\prime}
     \text{Tr}[H_{\nu}^{(1),\mu}(\xi,u,u^{\prime})
      \phi_{B}(\xi)
       \phi_{\pi}^{\nu}(u,u^{\prime})]\\
&=  & \int d\xi
     \int d u^{\prime}
      \text{Tr}[H^{(1),\mu}(\xi,u^{\prime})
       \phi_{B}(\xi)\phi_{\pi,n\cdot A}(u^{\prime})]
       - \int d\xi
          \int d u
           \text{Tr}[H^{(1),\mu}(\xi,u)\phi_{B}(\xi)\phi_{\pi,n\cdot A}(u)]\nonumber\\
&&+O(A_{\perp})
\,,\nonumber
\end{eqnarray}
where 
\begin{eqnarray}
\phi_{\pi,n\cdot A}(u) 
& = & 
\int\frac{d\lambda}{(2\pi)}
  e^{i\lambda u}
  \langle\pi|
     \bar{q}
       (\frac{\lambda}{E}n)
       (-i g)
          \int_{0}^{\infty}d\eta 
             n\cdot A^{a}(\frac{\eta}{E}n)
              T^{a}q(0)|0\rangle
\,.
\end{eqnarray}
$O(A_{\perp})$ denote the terms composed of gauge fields with a transversal
polarization and are identified as sub-leading twist contributions.
The gauge phase factors are absorbed into the first term in Eq.(\ref{eq:leading-amplitude}).
In covariant gauge, there are infinite number of similar gauge phase
factor terms from higher order Feynman diagrams. Their treatments
are similar to the above and skipped here. 
The result becomes 
\begin{eqnarray}
M^{\mu} 
& = & 
 \int d\xi
  \int d u
  \text{Tr}[H^{(1),\mu}(\xi,u)
   \phi_{B}(\xi)\phi_{\pi}(u)] \\
&  & 
   +   \int d\xi
         \int d u
          \int d u^{\prime}
           \text{Tr}[H_{\nu\eta}^{(1),\mu}(\xi,u,u^{\prime})
            \phi_{B}(\xi)w_{\eta^{\prime}}^{\eta} 
             \phi_{\pi}^{\eta^{\prime}\nu}(u,u^{\prime})]
            +\cdots
\,.
\nonumber
\end{eqnarray}

The color index factorization are performed in the following way,
\begin{eqnarray}
\text{Tr}_{c}
[H^{(1),\mu}(\xi,u)
 \phi_{B}(\xi)
  \phi_{\pi}(u)] 
& = & 
   [H^{(1),\mu}(\xi,u) ]_{ij,kl}
   [\phi_{B}(\xi) ]_{ik}
   [\phi_{\pi}(u) ]_{jl}
\,,\label{eq:colorfactor000-1}\\
\text{Tr}_{c}
 [H_{\nu\eta}^{(1),\mu }(\xi,u,u^{\prime})
  \phi_{B}(\xi )
   w_{\eta^{\prime}}^{\eta}
   \phi_{\pi}^{\eta^{\prime}\nu}(u,u^{\prime})
   ] 
& = & [ H_{\nu\eta}^{(1),\mu,b}(\xi,u,u^{\prime})
       ]_{ij,km}
      [\phi_{B}(\xi) ]_{ik}
      w_{\eta^{\prime}}^{\eta}
       [\phi_{\pi}^{\eta^{\prime}\nu, b}(u)]_{jm}
\,,\label{eq:colorfactor000-2}
\end{eqnarray}
where $i,\, j,\, k,\, l$ 
are color indices in the fundamental representation
and $b$ is the color index in the adjoint presentation. 
$[H^{(1),\mu}(\xi,u)]_{ij,kl}$ 
and 
$[H_{\nu\eta}^{(1),\mu,b}(\xi,u,u^{\prime})]_{ij,kl}$ 
are expanded in terms of color factors 
\begin{eqnarray}
[H^{(1),\mu}(\xi,u) ]_{ij,kl} 
& = & 
  H^{(1),\mu}(\xi,u)
   \left(
    \frac{1}{N_{c}^{2}}
    \delta_{ij}
     \delta_{kl}
    + (T^{a})_{ij}
      (T^{a})_{kl}
    + \cdots 
    \right )
\,,\label{eq:colorfactor-001}
\end{eqnarray}
\begin{eqnarray}
[
 H_{\nu\eta}^{(1),\mu,b}(\xi,u,u^{\prime})
 ]_{i j,k m} 
& = & 
   H_{\nu\eta}^{(1),\mu}(\xi,u,u^{\prime})
    \left(
     \frac{1}{N_{c}^{2}}
     \delta_{ij}
     \delta_{kl}
      (T^{b})_{lm}
    + (T^{a})_{ij}
      (T^{a})_{kl}
       (T^{b})_{lm}
    + \cdots 
      \right )
\,.\label{eq:colorfactor-002}
\end{eqnarray}
For $O(\alpha_{s})$ Feynman diagrams depicted in Fig.~\ref{fig:two parton}, only the
second terms in the right hand side of Eq.(\ref{eq:colorfactor-001},\ref{eq:colorfactor-002})
can contribute. They can be simplified by the color algebra
\begin{eqnarray}
(T^{a})_{ij}
(T^{a})_{kl} 
& = & 
\delta_{ik}
\delta_{jl}
\frac{1}{N_{c}^{2}}
\text{Tr}[T^{a}T^{a}]
=
\delta_{ik}
\delta_{jl}
\frac{C_{F}}{N_{c}}
\,,\label{eq:coloralgebra001}\\
(T^{a})_{ij}
(T^{a})_{kl}
(T^{b})_{lm} 
& = & 
\delta_{ik}
\delta_{jl}
\frac{1}{N_{c}^{2}}
\text{Tr}[T^{a}T^{a}]
(T^{b})_{lm}
=
\delta_{ik}
\delta_{jl}
\frac{C_{F}}{N_{c}}
(T^{b})_{lm}
\,.\label{eq:coloralgebra002}
\end{eqnarray}
Eq.(\ref{eq:coloralgebra001}) is applied to Eq.(\ref{eq:colorfactor000-1})
to have 
\begin{eqnarray}
\delta_{ik}
\delta_{jl}
\frac{C_{F}}{N_{c}}
[\phi_{B}(\xi)]_{ik}[\phi(u)]_{jl} 
& = & 
\frac{C_{F}}{N_{c}}
\phi_{B}(\xi)\phi_{\pi}(u)\,.
\end{eqnarray}
Similarly, Eq.(\ref{eq:coloralgebra002}) is applied to Eq.(\ref{eq:colorfactor000-2})
to have \begin{eqnarray}
\delta_{ik}
\delta_{jl}
\frac{C_{F}}{N_{c}}
(T^{b})_{lm}
[\phi_{B}(\xi)]_{ik}
(w_{\eta^{\prime}}^{\eta})
[\phi_{\pi}^{\eta^{\prime}\nu,b}(u,u^{\prime})]_{j m} 
& = & 
\frac{C_{F}}{N_{c}}
\phi_{B}(\xi)
(T^{b}\phi_{\pi}^{\eta^{\prime}\nu,b}(u,u^{\prime}))\,.
\end{eqnarray}

Before the spin index factorization is performed, it is necessary
to eliminate all terms in the $H$ functions which may lead to higher
twist contributions under the equation of motion for the quark. 
The equations of motion of light quarks (assumed mass-less), whose momentum
is $l^{\mu}$ in the collinear region 
$l^{\mu}=(l^{+},l^{-},l_{\perp})\sim(E,\Lambda/E,\Lambda)$,
are equivalently to the following identities
\begin{eqnarray}
\frac{i\not\! l_{L}}{l^{2}+i\epsilon}\not\!\bar{n} 
& = & 
   \frac{i\not\, l}{l^{2}+i\epsilon}
   (l-\hat{l})^{\alpha}
    (i\gamma_{\alpha})
     \frac{i\not\! n}{2n\cdot l+i\epsilon}
    \not\!\bar{n} \nonumber\\
& = & \omega_{\alpha^{\prime}}^{\alpha}
      \left[
       \frac{i\not\! l}{l^{2}+i\epsilon}
        l^{\alpha^{\prime}}
        \right]
       \left[
        (i\gamma_{\alpha})
        \frac{i\not\! n}{2n\cdot l+i\epsilon}
        \not\!\bar{n}
       \right]
\,.\label{eq:long-quark-prop}
\end{eqnarray}
Since the propagator (the special propagator) 
\[
\frac{i\not\! n}{2n\cdot l+i\epsilon}
\]
does not propagate, its associated terms are absorbed into the hard
scattering functions (the last square bracket term in the last term
of Eq.(\ref{eq:long-quark-prop})). 
The meanings of the above identities
are as follows. If there is one factor $\not\!\bar{n}$ in the hard
scattering functions contact with the long distance part of the quark
parton propagator of the $\pi$ meson, then the result is to extract
one short distance part of the propagator and a vertex $i\gamma_{\alpha}$
with a non-collinear momentum factor $(l-\hat{l})^{\alpha}$. 
The non-collinear momentum factor will be absorbed by the $\pi$ meson
function, $\phi_{\pi}(u)$, to have 
\begin{eqnarray}
(l-\hat{l})^{\alpha}\phi_{\pi}(u) 
& = & \omega_{\alpha^{\prime}}^{\alpha}
       \int\frac{d\lambda}{2\pi}
        \int\frac{d\eta}{2\pi}
        e^{i\lambda u}
        e^{i\eta(u^{\prime}-u)}
        \langle\pi|\bar{q}(\frac{n}{E}\lambda)
           i\partial^{\alpha^{\prime}}(\frac{\eta}{E}n)q(0)|0\rangle\\
 & = & \omega_{\alpha^{\prime}}^{\alpha}\phi_{\pi,\partial}^{\alpha^{\prime}}(u,u^{\prime})\,.
\end{eqnarray}
The similar case arises when the $\not\!\bar{n}$ in the hard scattering
functions contact with the short distance part of the quark parton
propagator of the $\pi$ meson, the result is 
\begin{eqnarray}
\frac{i\not\! n}{2 n\cdot l+i\epsilon}
 \not\!\bar{n} 
& = & 
\frac{i\not\! l^{\prime}}{l^{\prime2}+i\epsilon}
  (-gA^{\alpha})
  (i\gamma_{\alpha})\frac{i\not\! n}{2n\cdot l+i\epsilon}
   \not\!\bar{n} 
\\
& = & \omega_{\alpha^{\prime}}^{\alpha}
  \left[
    \frac{i\not\! l^{\prime}}{l^{\prime2}+i\epsilon}
      (-gA^{\alpha^{\prime}})
     \right]
     \left[
       (i\gamma_{\alpha})
        \frac{i\not\! n}{2n\cdot l+i\epsilon}
        \not\!\bar{n}
       \right]
    +\cdots
\,,\nonumber
\end{eqnarray}
where dots denotes the term would be absorbed by the gauge phase factor
of the matrix element. The $A^{\alpha}$ is the gauge field and its
associated terms are defined to be absorbed into their corresponding
hadron functions
\begin{eqnarray*}
\omega_{\alpha^{\prime}}^{\alpha}(-g A^{\alpha})\phi_{\pi}(u) 
& = & 
\omega_{\alpha^{\prime}}^{\alpha}
 \int\frac{d\lambda}{2\pi}
  \int\frac{d\eta}{2\pi}
   e^{i\lambda u}
    e^{i\eta(u^{\prime}-u)}
    \langle\pi|\bar{q}(\frac{n}{E}\lambda)
        (-g)A^{\alpha}(\frac{\eta}{E}n)q(0)|0\rangle \\
 & = & \omega_{\alpha^{\prime}}^{\alpha}
       \phi_{\pi,A}^{\alpha^{\prime}}(u,u^{\prime})\,. \nonumber
\end{eqnarray*}
The total effect, when one $\not\!\bar{n}$ factor can contact with
the partons of the $\pi$ meson, is 
\begin{eqnarray}
& & \text{\text{Tr}}\left[H_{\mu}^{(1)}(\xi,\pi)\phi_{B}(\xi)\phi_{\pi}(u)\right]\\
&= & \text{\text{Tr}}
\left[
  \left(
   (i\gamma_{\alpha})
   \frac{i\not\! n}{2n\cdot l_{q}+i\epsilon}
   H_{\mu}^{(1)}(\xi,u)
   +H_{\mu}^{(1)}(\xi,u)
    \frac{-i\not\! n}{2n\cdot l_{\bar{q}}-i\epsilon}
    (-i\gamma_{\alpha})
    \right)
    \omega_{\alpha^{\prime}}^{\alpha}
    \phi_{B}(\xi)
    \phi_{\pi,D}^{\alpha^{\prime}}(u)
    \right]
\;,
\nonumber
\end{eqnarray}
where 
\begin{eqnarray}
\phi_{\pi,D}^{\alpha^{\prime}}(u,u^{\prime}) 
& = & 
\int\frac{d\lambda}{2\pi}
 \int\frac{d\eta}{2\pi}
  e^{i\lambda u}
   e^{i\eta(u^{\prime}-u)}
   \langle\pi|\bar{q}(\frac{n}{E}\lambda)iD^{\alpha^{\prime}}(\frac{\eta}{E}n)q(0)|0\rangle\,,
\end{eqnarray}
where $iD^{\alpha^{\prime}}=i\partial^{\alpha^{\prime}}-gA^{\alpha^{\prime}}$.
Since $n\cdot l_{q(\bar{q})}$ are of order $E$ for collinear $l_{q(\bar{q})}$,
the related contributions are suppressed by one additional $E^{-1}$order.
We call these terms as \emph{abnormal} terms. 
The other terms are identified as \emph{normal} terms. 
The \emph{normal} terms will be kept in the reduced hard scattering functions, 
while the \emph{abnormal} terms are dropped.

In the form factors, the $\xi$ is of order $O(\Lambda/E)$ due to
the twist-2 $B$ meson distribution amplitude, $\phi_{B}^{tw2}(\xi)$.
The expansion into short distance and long parts of the quark propagator
can not help to separate different twist contributions from the $B$
meson. In this work, we only consider the contributions from the two
parton Fock state $|b\bar{q}\rangle$ of the $B$ meson. The contributions
from the sub-leading twist state of $B$ meson are left to other places.
The equation of motion for the $b$ quark, $(\not\! P_{b}-m_{b})b=0$,
is the only condition. This fact reflects in the derivation of the
reduced hard scattering functions. (Refer to following text and Appendix
A.) 

Considering all possibilities of the applications of equations of
motion of quarks, the spin structures of the hard scattering functions
are strongly restricted. It is useful to take the diagram in Fig.1(a)
as an example to explain this operation. After the color index factorization,
the amplitude for Fig.1(a) is proportional to 
\begin{equation}
\int d\xi
 \int d u
  \text{Tr}[H^{(1),\mu,(1a)}(\xi,u)\phi_{B}(\xi)\phi_{\pi}(u)]
\,.
\end{equation}
The spin structure of $H^{(1),\mu,(1a)}(\xi,u)$ has the expression
\begin{eqnarray}
[H^{(1),\mu,(1a)}(\xi,u)]_{ij,kl} 
& \propto & 
[\gamma^{\alpha}(\not\!\hat{l}_{q}-\not\! k)\gamma^{\mu}]_{ij}
 [\gamma_{\alpha}]_{kl}\,,
\end{eqnarray}
where $\not\!\hat{l}_{q}=u\not\! p_{\pi}$ and 
$\not\! k=\not\! n\bar{n}\cdot P_{b}-\bar{u}\not\! p_{\pi}$.
By using 
\begin{eqnarray}
g^{\alpha\beta} 
& = & 
\bar{n}^{\alpha}n^{\beta}
+ n^{\alpha}\bar{n}^{\beta}
+ d_{\perp}^{\alpha\beta}
\,,\\
\gamma^{\alpha} 
& = & 
\bar{n}^{\alpha}
\not\! n
+
n^{\alpha}\not\!\bar{n}
+
\gamma_{\perp}^{\alpha}
\,,
\end{eqnarray}
where 
$\gamma_{\perp}^{\alpha}
=d_{\perp}^{\alpha\beta}\gamma_{\beta}$
and 
$g_{\alpha}^{\alpha}=d_{\alpha}^{\alpha}=-2$. 
The spin structure of 
$H^{(1),\mu,(1a)}(\xi,u)$ 
is then expanded into a series in terms of 
$\not\!\bar{n}$, 
$\not\! n$, 
$\gamma_{\perp}^{\alpha}$. 
Each term in the expansion series is then examined to determine whether
it is of the considered twist or of higher twist according to the
equations of motion for the quarks in the $\pi$ meson and the $B$
meson. 
The abnormal terms are subtracted from the expression. 
We note that this analysis is automatically fulfilled for leading twist hard scattering functions,
because the leading twist spin structure of the $\pi$ meson can eliminate
the possible $\not\!\bar{n}$ factor by 
$\not\! p_{\pi}\not\!\bar{n}\propto\not\!\bar{n}\not\!\bar{n}=0$.
For the sub-leading twist amplitudes, 
this procedure of subtraction of possible abnormal terms is necessary. 
For example, the spin structures of the twist-3 PS or PT DA are proportional to $\gamma_{5}$ or
$\epsilon_{\mu\nu\alpha\beta}\sigma^{\mu\nu}\bar{n}^{\alpha}n^{\beta}$.
These spin factors can not eliminate the abnormal terms by the traditional method.
In the following, we assume that the hard scattering functions are
determined according to the above analysis 
and the resultant hard scattering functions are called reduced hard scattering functions. 
The relevant reduced hard scattering functions are given in Appendix A. 

The spin index factorization is performed in the following way, 
\begin{eqnarray}
\text{Tr}_{s}[H^{(1),\mu}(\xi,u)\phi_{B}(\xi)\phi_{\pi}(u)] 
& = & 
 [H^{(1),\mu}(\xi,u) ]_{i j,k l}
 [\phi_{B}(\xi) ]_{i k}
 [\phi_{\pi}(u)]_{j l}
\,,\label{eq:spinfactor000-1}\\
\text{Tr}_{s}
 [H_{\nu\eta}^{(1),\mu}(\xi,u,u^{\prime})
  \phi_{B}(\xi)
   w_{\eta^{\prime}}^{\eta}
   \phi_{\pi}^{\eta^{\prime}\nu}(u,u^{\prime})] 
& = & 
  [H_{\nu\eta}^{(1),\mu}(\xi,u,u^{\prime})]_{i j,kl}
  [\phi_{B}(\xi)]_{i k}
  w_{\eta^{\prime}}^{\eta}
  [\phi_{\pi}^{\eta^{\prime}\nu}(u)]_{j l}
\,,  
\label{eq:spinfactor000-2}
\end{eqnarray}
where the meson functions are expanded as 
\begin{eqnarray}
[\phi_{B}(\xi) ]_{i k} 
 & = & \frac{1}{4}(\gamma^{\rho}\gamma_{5})_{i k}
       \text{Tr}[\phi_{B}(\xi)\gamma_{\rho}\gamma_{5}]
      +\frac{1}{4}(\gamma_{5})_{i k}
        \text{Tr}[\phi_{B}(\xi)\gamma_{5}]
      +\cdots 
\,,\\
\left[
 \phi_{\pi}(u)
 \right ]_{j l} 
& = & \frac{1}{4}(\gamma^{\rho}\gamma_{5})_{j l}
       \text{Tr}[\phi_{\pi}(u)\gamma_{\rho}\gamma_{5}]
      +\frac{1}{4}(\gamma_{5})_{j l}
        \text{Tr}[\phi_{\pi}(\xi)\gamma_{5}]
\\        
&  &    +\frac{1}{8}(\sigma_{\rho\lambda}\gamma_{5})_{j l}
        \text{Tr}[\phi_{\pi}(u)\sigma^{\rho\lambda}\gamma_{5}]
      +\cdots
\,,\nonumber\\
\left[
\phi_{\pi}^{\eta^{\prime}\nu}(u,u^{\prime})
\right ]_{j l} 
& = & \frac{1}{8}(\sigma_{\rho\lambda}\gamma_{5})_{j l}
       \text{Tr}[\phi_{\pi}^{\eta^{\prime}\nu}(u,u^{\prime})
                  \sigma^{\rho\lambda}\gamma_{5}]
      + \cdots
\,.
\end{eqnarray}
The dots denote those terms are not of our interesting. 
Each coefficient in the above expansion is attributed by a DA according to the following
definitions 
\begin{eqnarray}
\text{Tr}[\phi_{B}(\xi)\gamma_{\rho}\gamma_{5}] 
& = & if_{B}[P_{B,\rho}\phi_{B}(\xi)
     + E(n_{\rho}-\bar{n}_{\rho})
       \bar{\phi}_{B}]
     +\cdots
\,,\\
\text{Tr}[\phi_{B}(\xi)\gamma_{5}] 
& = & if_{B}m_{B}\phi_{B}(\xi)
      +\cdots
\,,\\
\text{Tr}[\phi_{\pi}(u)\gamma_{\rho}\gamma_{5}] 
& = & -if_{\pi}p_{\pi,\rho}\phi_{\pi}^{P}(u)
      +\cdots
\,,\\
\text{Tr}[\phi_{\pi}(\xi)\gamma_{5}] 
& = & -if_{\pi}\mu_{\chi}\phi_{\pi}^{p}(u)
      +\cdots
\,,\\
\text{Tr}[\phi_{\pi}(u)\sigma^{\rho\lambda}\gamma_{5}]
& = & -f_{\pi}\mu_{\chi}
      [\bar{n}^{\rho},n^{\lambda}]
      \phi_{\pi}^{\sigma}(u)
      +\cdots
\,,\\
\text{Tr}[\phi_{\pi}^{\eta^{\prime}\nu}(u,u^{\prime})\sigma^{\rho\lambda}\gamma_{5}] 
& = & -f_{\pi}\mu_{\chi}
       (p_{\pi}^{\rho}d^{\lambda\eta^{\prime}}-p_{\pi}^{\lambda}d^{\rho\eta^{\prime}})p_{\pi}^{\nu}
        \phi_{\pi}^{3p}(u,u^{\prime})
       +\cdots
\,.
\end{eqnarray}
In these definitions for the DAs, only relevant ones are shown and
the others are left into the dots terms. 
Note that the $\phi_{\pi}^{p,\sigma}(u)$ are defined as 
the the energetic limits of the PS and PT DAs of the
$\pi$ meson, respectively. 
The spin projector of $\phi_{\pi}^{\sigma}(u)$
is the leading part of the full spin projector $[p_{\pi}^{\rho},z^{\lambda}]$
of the PT DA under the energetic limit. 
The $\phi_{\pi}^{\sigma}(u)$
corresponds to the $d\phi_{\pi}^{\sigma,c}(u)/du$ of the usually
used PT DA, $\phi_{\pi}^{\sigma,c}(u)$. 
The factor $\mu_{\chi}=m_{\pi}^{2}/(m_{q}+m_{\bar{q}})$
with $m_{\pi}$ the pion mass and $m_{q(\bar{q})}$ the current quark(anti-quark) masses. 
The Dirac matrices in the expansion series are absorbed by the hard scattering functions. 
The spin index factorization is completed.

The final result is written as 
\begin{eqnarray}
M^{\mu} 
& = & M^{tw2,\mu}
     + M^{B,\mu}
     + M^{tw3,ps,\mu}
     + M^{tw3,pt,\mu}
     + M^{tw3,3p,\mu}
     + \cdots\,,
     \label{eq:amplitudeexpan}
\end{eqnarray}
where the expansion terms in the right hand side of Eq.(\ref{eq:amplitudeexpan})
are defined as 
\begin{eqnarray}
M^{tw2,\mu} 
& = & \frac{1}{16}
       f_{B}f_{\pi}
       \int d\xi \phi_{B}(\xi)
       \int d u \phi_{\pi}^{P}(u)
        \text{Tr}
        \left[
         \gamma_{5}
          \not\! p_{\pi} 
          H^{(1),\mu}(\xi,u)
          (\not\! P_{B}+m_{B})
           \gamma_{5}
           \right]\,,\\
M^{B,\mu} 
& = & \frac{1}{16} 
       f_{B}f_{\pi} E
       \int d\xi \bar{\phi}_{B}(\xi)
       \int d u\phi_{\pi}^{P}(u)
       \text{Tr}
        \left[
         \gamma_{5}
          \not\! p_{\pi}
          H^{(1),\mu}(\xi,u)
          (\not\! n-\not\!\bar{n})
           \gamma_{5}
          \right]\,,\\
M^{tw3,ps,\mu} 
& = & \frac{1}{16}
      f_{B}f_{\pi}
       \mu_{\chi}
       \int d\xi \phi_{B}(\xi)
       \int d u\phi_{\pi}^{p}(u)
       \text{Tr}
        \left[
         \gamma_{5}
          H^{(1),\mu}(\xi,u)
          (\not\! P_{B}+m_{B})
          \gamma_{5}
           \right]\,,
           \label{eq:amp-tw3-ps}\\
M^{tw3,pt,\mu} 
& = & -\frac{i}{32}
       f_{B}f_{\pi}
        \mu_{\chi}
        \int d\xi \phi_{B}(\xi)
         \int d u\phi_{\pi}^{\sigma}(u)
         \text{Tr}
          \left[
           \epsilon_{\perp}\cdot\sigma
            H^{(1),\mu}(\xi,u)
           (\not\! P_{B}+m_{B})
            \gamma_{5}
            \right]\,,
             \label{eq:amp-tw3-pt}\\
M^{tw3,3p,\mu} 
& = & -\frac{i}{32}
        f_{B}f_{\pi}
        \mu_{\chi}
        \int d\xi \phi_{B}(\xi)
        \int d u_{q} 
         \int d u_{\bar{q}}
          \int d u_{g}           
            \phi_{\pi}^{3 p} (u_{q},u_{\bar{q}},u_{g})
\\
\label{eq:amp-tw3-3p} 
&  & \times              
      \delta(1-u_{q}-u_{\bar{q}}-u_{g})
            \text{Tr}
             \left[
               \sigma_{\rho\lambda}\gamma_{5}
                H_{\eta\nu}^{(1),\mu}(\xi,u_{q},u_{\bar{q}},u_{g})
                (\not\! P_{B}+m_{B})
                 \gamma_{5}
                  \right]
                   \omega_{\eta^{\prime}}^{\eta}
                   \Gamma^{\rho\lambda\eta^{\prime}\nu}(p_{\pi})\,,
\nonumber 
\end{eqnarray}
where 
$
\Gamma^{\rho\lambda\eta^{\prime}\nu}(p_{\pi})
=(p_{\pi}^{\rho}d^{\lambda\eta^{\prime}}
  -p_{\pi}^{\lambda}d^{\rho\eta^{\prime}})
   p_{\pi}^{\nu}
$
and 
$
\epsilon_{\perp}\cdot\sigma
=\epsilon_{\perp,\alpha\beta}
  \sigma^{\alpha\beta}
= \epsilon_{\alpha\beta\eta\gamma}
  \sigma^{\alpha\beta}\bar{n}^{\eta}n^{\gamma}
$
with 
$\sigma^{\alpha\beta}
=i[\gamma^{\alpha},\gamma^{\beta}]/2
$.
$M^{tw2,\mu}$ has been calculated in previous section. 
The other four twist-3 terms, $M^{tw3,B,ps,pt,3p,\mu}$, are calculated in next
Section. 
In the above expression, we have used a transformation of
the integrations over $u$ and $u^{\prime}$ into the integrations
over $u_{q}$, $u_{\bar{q}}$, and $u_{g}$ to make the expression
more explicitly. 
The $u_{q}$, $u_{\bar{q}}$, and $u_{g}$ are defined
as the fractions of the quark, anti-quark, and gluon partons of the
$\pi$ meson. 

The retain of $\xi$ in the internal $b$ quark propagator can partially
resolve the end-point divergences at twist-3, which are linear divergences
due to the constant nature of the $\pi$ meson's twist-3 distribution
amplitude (the pseudo-scalar one, $\phi_{\pi}^{p}(u)$). The PS DA
$\phi_{\pi}^{p}(u)=1$ for the $\pi$ meson is determined by the equation
of motion for the quark of the $\pi$ meson in the soft pion state,
where the energy of the pion, $E_{\pi}$, the pion mass, $m_{\pi}$,
or the scale of the parton transverse momentum, $l_{\perp}$, are
of the same order, $E_{\pi}\simeq m_{\pi}\simeq l_{\perp}\simeq O(\Lambda)$
. The final state $\pi$ meson in the semi-leptonic $\bar{B^{0}}\to\pi^{-}l^{+}\nu_{l}$
process can carry an energetic momentum, where $E_{\pi}\gg m_{\pi}$.
The equation of motion for $\phi_{\pi}^{p}(u)$ can be further approximated
to a reduced equation of motion \cite{Yeh:2007fi}. According to this
reduced equation of motion, the $\pi$ meson's PS DA and the PT DA,
$\phi_{\pi}^{\sigma}(u)$, become equal. In this way, $\phi_{\pi}^{p}(u)$
and $\phi_{\pi}^{\sigma}(u)$ can be modeled similar to the twist-2
$\pi$ meson's distribution amplitude, $\phi_{\pi}^{P}(u)$, to take
their asymptotic forms, $\phi_{\pi}^{p}(u)=\phi_{\pi}^{\sigma}(u)=6u(1-u)$.
The constant feature of $\phi_{\pi}^{p}(u)$ as found in literature
is disappeared in the $\pi$ meson's energetic state. 

In the $\pi$ meson's soft pion state, different twist order contributions
ordered by $O((\Lambda/E_{\pi})^{n})$ become equally important under
the limit $E_{\pi}\simeq\Lambda$. 
On the other hand, in the $\pi$ meson's energetic state \cite{Yeh:2007fi}, 
different twist order contributions ordered by $O((\Lambda/E_{\pi})^{n})$ are restrictively ordered under
the limit $E_{\pi}\gg\Lambda$. 
This explains why $\phi_{\pi}^{p}(u)$
can have different forms in the energetic and chiral limits. Once
the non-constant $\phi_{\pi}^{p}(u)$ are substituted into the relevant
expressions for the amplitude under the $\xi^{R}$, the linear divergences
automatically disappear and the related end-point divergent problem
is resolved. There are no similar end-point divergences for the twist-3
three parton amplitude $M^{tw3,3p,\mu}$ since the three parton DA
$\phi_{\pi}^{3p}(u_{q},u_{\bar{q}},u_{g})$ is not a constant. As
a result, we obtain a factorization formula for the $B\to\pi$ form
factors up-to twist-3 order.

\begin{figure}[t]
\includegraphics[scale=0.6]{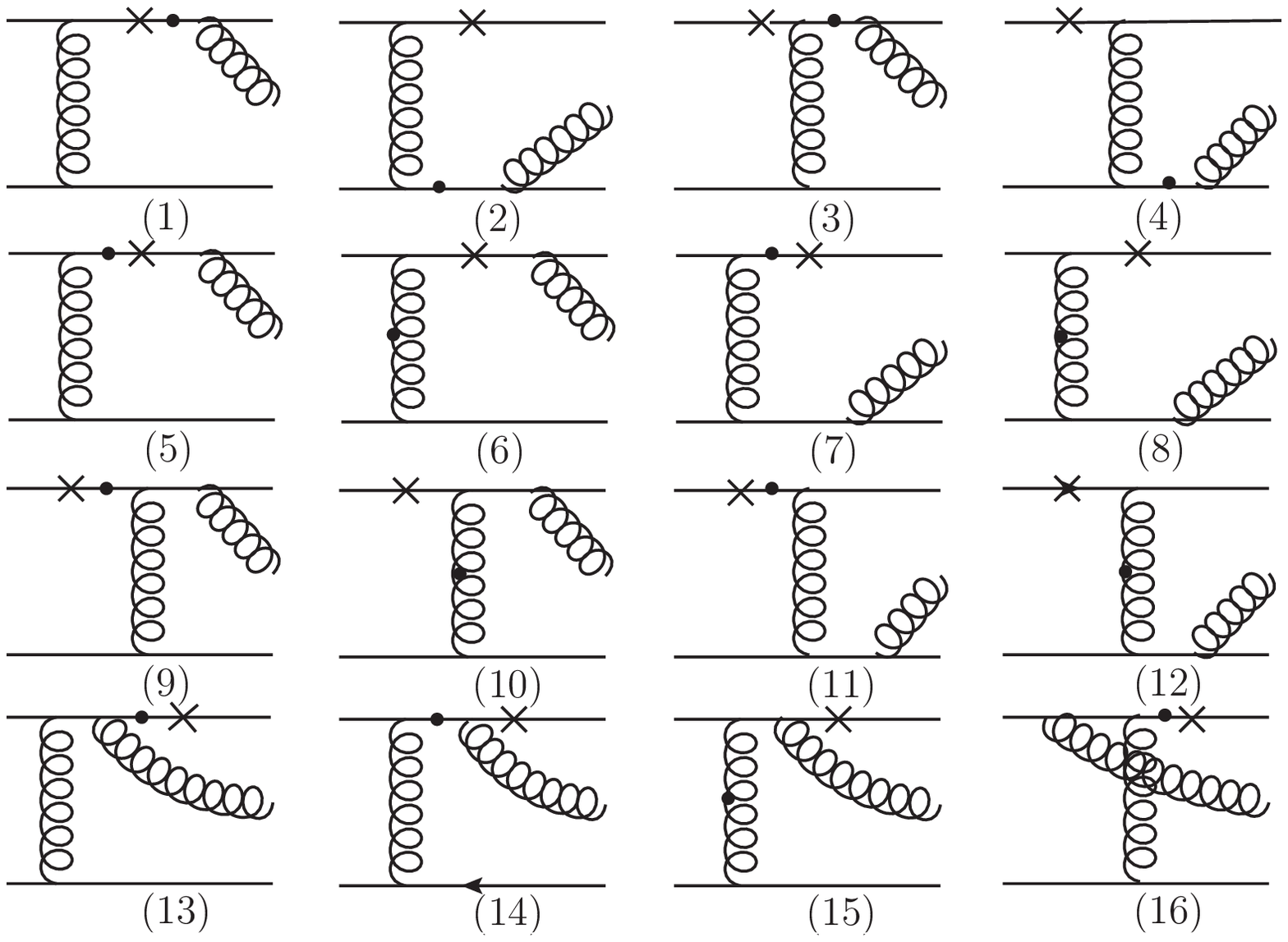}
\includegraphics[scale=0.6]{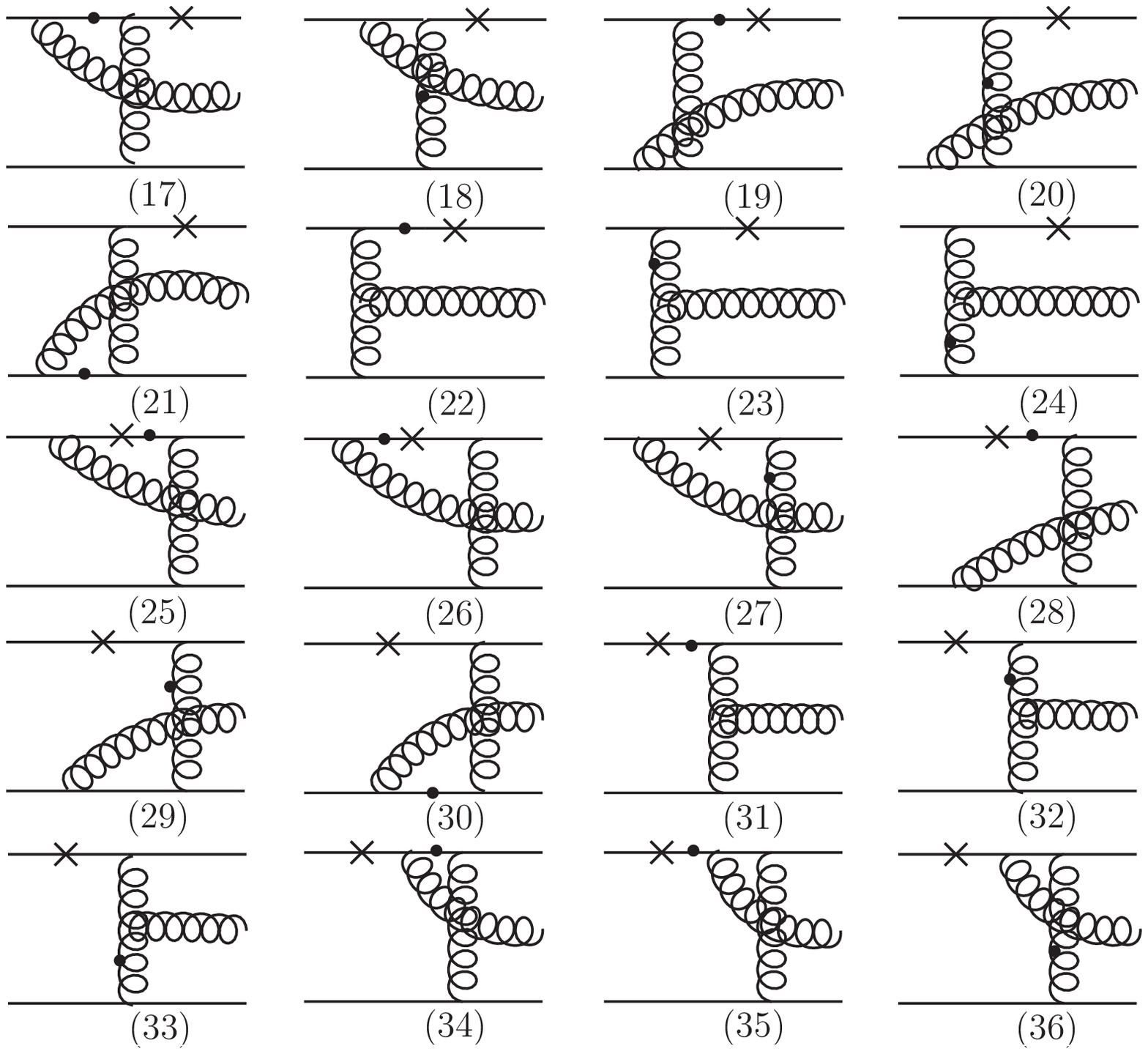}
\caption{Feynman diagrams for the hard scattering functions $H^{\mu}_{\nu\eta}$ of the 
$C^f$ amplitudes for the $\bar{B}\to \pi l\bar{\nu}_l$ decays. 
The external meson states $|\bar{B}\rangle$ and $\langle\pi|$ are not shown.
The cross vertex denotes the vector operator $\gamma^\mu$.
The dot vertex denotes the subscript $\eta$ in $H^{\mu}_{\nu\eta}$.
Only contributions from diagrams (1-4) are of twist-3 order.
The contributions from diagram (5-36) are of higher than twist-3 order.}
\label{fig:fig3}
\end{figure}
\newpage
The calculations of three parton contributions are straightforward
by using the reduced hard scattering functions given in Appendix A.
The important steps in the calculations are described below. There
are totally 36 Feynman diagrams as depicted in Fig.~\ref{fig:fig3}. Only 4 Feynman
diagrams (Fig.~\ref{fig:fig3}(1)-(4)) are needed at twist-3 order. 
The rest 32 Feynman diagrams (Fig.~\ref{fig:fig3}(5)-(36)) are of higher twist. 
Under covariant gauge, the contraction of the spin projector
$
\Gamma^{\rho\lambda\eta\nu}(p_{\pi})
=(p_{\pi}^{\rho}d^{\lambda\eta}-p_{\pi}^{\lambda}d^{\rho\eta})p_{\pi}^{\nu}
$
with those contributions from Fig.~\ref{fig:fig3}(1)-(4) contained in 
$
\text{Tr}
 \left[
     \sigma_{\rho\lambda}\gamma_{5} 
     H_{\eta\nu}^{(1),\mu}(\xi,u_{q},u_{\bar{q}},u_{g})
     (\not\! P_{B}+m_{B})
     \gamma_{5}
  \right]
$
results in Eq.(\ref{eq:hardfunct-tw3-3p}).

\section{Twist-3 amplitudes and form factors}

In this Section, we present the twist-3 amplitudes 
$M^{B,\mu}$, $M^{tw3,ps,\mu}$,
$M^{tw3,pt,\mu}$, 
and $M^{tw3,3p,\mu}$ calculated by the collinear
expansion introduced in last Section. 
The twist-3 form factors $f_{1,2}^{tw3}$
are extracted from each twist-3 amplitude. 
The twist-3 form factors
$F_{+,0}^{B\pi,tw3}$ are determined from the twist-3 form factors
$f_{1,2}^{tw3}$. 
The calculations of 
$M_{\mu}^{tw3,ps}$, 
$M^{B,\mu}$,
$M^{tw3,ps,\mu}$, 
$M^{tw3,pt,\mu}$, 
and $M^{tw3,3p,\mu}$ are straightforward
according to 
Eqs.(\ref{eq:amp-tw3-ps},\ref{eq:amp-tw3-pt},\ref{eq:amp-tw3-3p}).
The reduced two parton hard scattering functions 
$H^{(1),\mu}(\xi,u)$
and the reduced three parton hard scattering functions $H_{\nu\eta}^{(1),\mu}$
are given in Appendix A. 
By substituting these reduced hard scattering
functions into Eqs.(\ref{eq:amp-tw3-ps},\ref{eq:amp-tw3-pt},\ref{eq:amp-tw3-3p}),
one can obtain the following twist-3 contributions.

\subsection{$\bar{\phi}_{B}$ contributions}

The $\bar{\phi}_{B}$ contributions are written as 
\begin{eqnarray}
M^{B,\mu} 
& = & 
  \frac{\pi\alpha_{s}C_{F}}{N_{c}}
   f_{\pi}f_{B}
    \int_{0}^{1}d\xi\bar{\phi}_{B}(\xi)
     \int_{0}^{1}d u\phi_{\pi}^{P}(u) H^{B,\mu}(\xi,u)
\,,     
\end{eqnarray}
where 
\begin{eqnarray*}
H^{B,\mu} 
& = & 
 \frac{1}{\eta\xi\bar{u}E^{2}} P_{B}^{\mu}
 +  \frac{(1+\eta)\eta\bar{u}-\eta-(1-\eta)\xi}{\eta^{2}\xi\bar{u}(\xi-\eta\bar{u})E^{2}}p_{\pi}^{\mu}
\,.
\end{eqnarray*}
Similar to the twist-2 case, the form factors are defined as 
\begin{eqnarray}
M^{B,\mu} 
& = & 
f_{1}^{B}(q^{2})P_{B}^{\mu}+f_{2}^{B}(q^{2})p_{\pi}^{\mu}
\,,
\end{eqnarray}
where $f_{1,2}^{B}$ are given by 
\begin{eqnarray}
f_{1}^{B}(q^{2}) 
& = & 
\frac{\pi\alpha_{s}}{E^{2}}
  \frac{C_{F}}{N_{c}}
  f_{\pi}f_{B}
  \int d\xi \bar{\phi}_{B}(\xi)
   \int d u \phi_{\pi}^{P}(u)
    \left[\frac{1}{\eta\xi\bar{u}}\right]
\,,\\
f_{2}^{B}(q^{2}) 
& = & 
\frac{\pi\alpha_{s}}{E^{2}}\frac{C_{F}}{N_{c}}
  f_{\pi}f_{B}
  \int d\xi\bar{\phi}_{B}(\xi)
   \int d u\phi_{\pi}^{P}(u)
    \left[
      \frac{(1+\eta)\eta\bar{u}-\eta-(1-\eta)\xi}{\eta^{2}\xi\bar{u}(\xi-\eta\bar{u})}
    \right]
\,.
\end{eqnarray}
The form factors $F_{+,0}^{B}(q^{2})$ are obtained by $f_{1,2}^{tw3,B}$ as 
\begin{eqnarray}
F_{+}^{B}(q^{2}) 
& = & 
\frac{\pi\alpha_{s}}{2 E^{2}}
 \frac{C_{F}}{N_{c}}
  f_{\pi}f_{B}
  \int d\xi\bar{\phi}_{B}(\xi)
   \int d u\phi_{\pi}^{P}(u)
    [\frac{\eta\bar{u}-\eta-(1-2\eta)\xi}{\eta^{2}\xi\bar{u}(\xi-\eta\bar{u})}]
\,,
\end{eqnarray}
\begin{eqnarray}
F_{0}^{B}(q^{2}) 
& = & 
\frac{\pi\alpha_{s}}{2 E^2}
 \frac{C_{F}}{N_{c}}
 f_{\pi}f_{B}
 \int d\xi\bar{\phi}_{B}(\xi)
  \int d u\phi_{\pi}^{P}(u)
   \left[
      \frac{\xi-(1-2\eta)\eta\bar{u}-\eta}{\eta\xi\bar{u}(\xi-\eta\bar{u})}
   \right]
\,.
\end{eqnarray}

\subsection{Twist-3 two parton pseudo-scalar and pseudo-tensor contributions}

The pseudo-scalar contributions and pseudo-tensor contributions are
found identical as 
\begin{eqnarray*}
M^{tw3,ps,\mu} 
& =M^{tw3,pt,\mu}= & 
f_{\pi}f_{B}
\frac{\pi\alpha_{s}C_{F}}{2N_{c}}
  r_{\chi}
   \int_{0}^{1}d\xi\phi_{B}(\xi)
    \int_{0}^{1}d u\phi_{\pi}^{p}(u)H^{tw3,2p,\mu}(\xi,u)
\,,
\end{eqnarray*}
where 
\begin{eqnarray*}
H^{tw3,2p,\mu} 
& = & 
-\frac{1}{\eta^{2}\bar{u}\xi E^{2}}P_{B}^{\mu}
 -\frac{(1+\eta)\eta\bar{u}-\xi}{\eta^{3}\xi\bar{u}(\xi-\eta\bar{u})E^{2}}p_{\pi}^{\mu}
\,.
\end{eqnarray*}
The form factors are defined as 
\begin{eqnarray}
M^{tw3,ps(pt),\mu} 
& = & 
  f_{1}^{tw3,ps(pt)}(q^{2})P_{B}^{\mu}
+ f_{2}^{tw3,ps(pt)}(q^{2})p_{\pi}^{\mu}
\;,
\end{eqnarray}
where the form factors 
$f_{1,2}^{tw3,ps(pt)}$ are given by 
($2E^{2}=m_{B}^{2}$
and $r_{\chi}=2\mu_{\chi}/m_{B}$) 
\begin{eqnarray}
f_{1}^{tw3,ps(pt)}(q^{2}) 
& = & 
-\frac{\pi\alpha_{s}}{2E^{2}}
  \frac{C_{F}}{N_{c}}
   f_{\pi}f_{B}
    r_{\chi}
    \int d\xi \phi_{B}(\xi)
     \int d u \phi_{\pi}^{p}(u)
     \left[
           \frac{1}{\eta^{2}\bar{u}\xi}
     \right]
\,,\\
f_{2}^{tw3,ps(pt)}(q^{2}) 
& = & 
- \frac{\pi\alpha_{s}}{2E^{2}}
   \frac{C_{F}}{N_{c}}
   f_{\pi}f_{B}
   r_{\chi}
   \int d\xi\phi_{B}(\xi)
    \int d u\phi_{\pi}^{p}(u)
    \left[
     \frac{(1+\eta)\eta\bar{u}-\xi}
          {\eta^{3}\xi\bar{u}(\xi-\eta\bar{u})}
     \right]
\,.
\end{eqnarray}
The form factors 
$F_{+,0}^{B\pi,tw3,ps}(q^{2})$ are obtained by 
$f_{1,2}^{tw3,ps}$ as 
\begin{eqnarray}
F_{+}^{B\pi,tw3,ps(pt)}(q^{2}) 
& = & 
- \frac{\pi\alpha_{s}}{4E^{2}}
   \frac{C_{F}}{N_{c}}
   f_{\pi}f_{B}
   r_{\chi}
   \int d\xi\phi_{B}(\xi)
   \int d u\phi_{\pi}^{p}(u)
    \left[
      \frac{\eta\bar{u}-(1-\eta)\xi}
           {\eta^{3}\xi\bar{u}(\xi-\eta\bar{u})}
     \right]
\,,
\end{eqnarray}
\begin{eqnarray}
F_{0}^{B\pi,tw3,ps(pt)}(q^{2}) 
& = & 
\frac{\pi\alpha_{s}}{4E}
  \frac{C_{F}}{N_{c}}
  f_{\pi}f_{B}
  r_{\chi}
   \int d\xi\phi_{B}(\xi)
    \int d u\phi_{\pi}^{p}(u)\\
&  & 
  \times
   \left[
     \frac{(1-2\eta)\eta\bar{u}-(1-\eta)\xi}{\eta^{2}\xi\bar{u}(\xi-\eta\bar{u})}
     \right]
\,.\nonumber 
\end{eqnarray}

We found that the twist-3 two pseudo-tensor amplitude $M^{tw3,pt,\mu}$
is equal to the twist-3 two pseudo-scalar amplitude $M^{tw3,ps,\mu}$.
The corresponding form factors defined by these two amplitudes are identical.
This is consistent with the reduced equations of motion 
$\phi_{\pi}^{p}(u)=\phi_{\pi}^{\sigma}(u)$.

\subsection{Twist-3 three parton contributions}

The twist-3 three parton contributions are written as 
\begin{eqnarray}
M^{tw3,3p,\mu} 
& = & 
\frac{\pi\alpha_{s}C_{F}}{N_{c}}
  f_{\pi}f_{B}
   r_{\chi}^{\pi}
   \int d\xi\phi_{B}(\xi)
    \int d u_{q} \int d u_{\bar{q}} \int d u_{g}
\\
 &  & 
   \times
    \delta(1-u_{q}-u_{\bar{q}}-u_{g})
    \phi_{\pi}^{3p}(u_{q},u_{\bar{q}},u_{g})
     H^{3p,\mu}(\xi,u_{q},u_{\bar{q}},u_{g})
\,,\nonumber 
\end{eqnarray}
where 
\begin{eqnarray}
H^{3p,\mu}(\xi,u_{q},u_{\bar{q}},u_{g}) 
& = & 
\frac{1}{u_{q}u_{\bar{q}}u_{g}}
\times
\left[
      \frac{1}{\xi\eta^{2}\bar{u}_{g}E^{2}}
         P_{B}^{\mu}
    + \frac{(1+\eta\bar{u}_{g})\eta u_{\bar{q}}-\xi}
         {\eta^{3}\xi(\xi-\eta u_{\bar{q}})E^{2}}
          p_{\pi}^{\mu}
 \right]
\,.\label{eq:hardfunct-tw3-3p}
\end{eqnarray}
The twist-3 three parton form factors are defined as 
\begin{eqnarray*}
M^{tw3,3p,\mu} 
& = & 
  f_{1}^{tw3,3p}(q^{2})P_{B}^{\mu}
+ f_{2}^{tw3,3p}(q^{2})p_{\pi}^{\mu}
\,,
\end{eqnarray*}
where the form factors $f_{1,2}^{tw3,3p}$ are given by 
($2E^{2}=m_{B}^{2}$
and $r_{\chi}=2\mu_{\chi}/m_{B}$) 
\begin{eqnarray}
f_{1}^{tw3,3p}(q^{2}) 
& = & 
 \frac{\pi\alpha_{s}}{E^{2}}
  \frac{C_{F}}{N_{c}}
   f_{\pi}f_{B}
   r_{\chi}
   \int d\xi\phi_{B}(\xi)
   \int d u_{q}\int d u_{\bar{q}}\int d u_{g}
    \delta(1-u_{q}-u_{\bar{q}}-u_{g})
\\
&  & 
  \times
  \frac{\phi_{\pi}^{3p}(u_{q},u_{\bar{q}},u_{g})}
       {u_{q}u_{\bar{q}}u_{g}}
  \left[
       \frac{1}{\eta^{2}\xi\bar{u}_{g}}
   \right]
\,,\nonumber \\
f_{2}^{tw3,3p}(q^{2}) 
& = & 
 \frac{\pi\alpha_{s}}{E^{2}}
  \frac{C_{F}}{N_{c}}
   f_{\pi}f_{B}r_{\chi}
   \int d\xi\phi_{B}(\xi)
    \int d u_{q}
     \int d u_{\bar{q}}
      \int d u_{g}
      \delta(1-u_{q}-u_{\bar{q}}-u_{g})
\\
&  & 
\times
   \frac{\phi_{\pi}^{3p}(u_{q},u_{\bar{q}},u_{g})}{u_{q}u_{\bar{q}}u_{g}}
   \left[
        \frac{(1+\eta\bar{u}_{g})\eta u_{\bar{q}}-\xi}
             {\eta^{3}\xi\bar{u}_{g}(\xi-\eta u_{\bar{q}})}
   \right]
\,.
\nonumber 
\end{eqnarray}
The form factors $F_{+,0}^{B\pi,tw3,3p}(q^{2})$ are obtained by 
Eq.(\ref{eq:f12toFFplus},\ref{eq:f12toFFzero})
\begin{eqnarray}
F_{+}^{B\pi,tw3,3p}(q^{2}) 
& = & 
\frac{\pi\alpha_{s}}{2E^{2}}
  \frac{C_{F}}{N_{c}}
   f_{\pi}f_{B}
   r_{\chi}^{\pi}
    \int d\xi\phi_{B}(\xi)
    \int d u_{q}\int d u_{\bar{q}}\int d u_{g}
     \delta(1-u_{q}-u_{\bar{q}}-u_{g})
\\
&  & 
\times
    \frac{\phi_{\pi}^{3p}(u_{q},u_{\bar{q}},u_{g})}{u_{q}u_{\bar{q}}u_{g}}
    \left[
         \frac{(1-\eta u_{g})\eta u_{\bar{q}}-\bar{\eta}\xi}
              {\eta^{3}\xi\bar{u}_{g}(\xi-\eta u_{\bar{q}})}
    \right]
\,,
\nonumber 
\end{eqnarray}
\begin{eqnarray}
F_{0}^{B\pi,tw3,3p}(q^{2}) 
& = & 
  \frac{\pi\alpha_{s}}{2E^{2}}
   \frac{C_{F}}{N_{c}}
   f_{\pi}f_{B}
   r_{\chi}^{\pi}
   \int d\xi\phi_{B}(\xi)
    \int d u_{q}\int d u_{\bar{q}}\int d u_{g}
     \delta(1-u_{q}-u_{\bar{q}}-u_{g})
\\
&  & 
\times
   \frac{\phi_{\pi}^{3p}(u_{q},u_{\bar{q}},u_{g})}
        {u_{q}u_{\bar{q}}u_{g}}
   \left[
         \frac{\bar{\eta}\xi-(1-\eta(2-u_{g}))\eta u_{\bar{q}}}
              {\eta^{2}\xi\bar{u}_{g}(\xi-\eta u_{\bar{q}})}
    \right]
\,.
\nonumber 
\end{eqnarray}

\subsection{The resultant form factors}

The form factors up-to $O(1/E)$ are 
\begin{eqnarray}
\label{F+total}
F_{+}^{B\pi}(q^{2}) 
& = & 
 F_{+}^{tw2}(q^{2})+F_{+}^{B}(q^{2})+F_{+}^{tw3,ps}(q^{2})+F_{+}^{tw3,pt}(q^{2})+F_{+}^{tw3,3p}(q^{2})
\\
& = & 
 \frac{\pi\alpha_{s}(t)}{2E^{2}}
 \frac{C_{F}}{N_{c}}
  f_{\pi}f_{B}
  \int d\xi\phi_{B}(\xi)
   \int du
   \left[
    \frac{\phi_{\pi}^{P}(u)[\eta-\xi+\eta\bar{u}]+\phi_{\pi}^{p}(u)r_{\chi}^{\pi}[(1-\eta)\xi-\eta\bar{u}]}
         {\eta^{2}\xi\bar{u}(\xi-\eta\bar{u})}
   \right]
\nonumber \\
&  & 
+ \frac{\pi\alpha_{s}(t)}{2E^{2}}
   \frac{C_{F}}{N_{c}}
    f_{\pi}f_{B}
     \int d\xi\bar{\phi}_{B}(\xi)
      \int d u\phi_{\pi}^{P}(u)
      \left[
        \frac{\eta\bar{u}-\eta-(1-2\eta)\xi}{\eta^{2}\xi\bar{u}(\xi-\eta\bar{u})}
      \right]
\nonumber \\
&  & 
+  \frac{\pi\alpha_{s}(t)}{2E^{2}}
    \frac{C_{F}}{N_{c}}
     f_{\pi}f_{B}
     r_{\chi}^{\pi}
     \int d\xi\phi_{B}(\xi)
     \int du_{q}\int du_{\bar{q}}\int du_{g}\delta(1-u_{q}-u_{\bar{q}}-u_{g})
\nonumber \\
&  & 
\times
  \frac{\phi_{\pi}^{3p}(u_{q},u_{\bar{q}},u_{g})}{u_{q}u_{\bar{q}}u_{g}}
 \left[
   \frac{(1-\eta u_{g})\eta u_{\bar{q}}-\bar{\eta}\xi}
        {\eta^{3}\xi\bar{u}_{g}(\xi-\eta u_{\bar{q}})\xi}
  \right]
\,,
\nonumber 
\end{eqnarray}
\begin{eqnarray}
F_{0}^{B\pi}(q^{2}) 
& = & 
F_{0}^{tw2}(q^{2})+F_{0}^{B}(q^{2})+ F_{0}^{tw3,ps}(q^{2}) + F_{0}^{tw3,pt}(q^{2}) + F_{0}^{tw3,3p}(q^{2})
\\
& = & 
 \frac{\pi\alpha_{s}(t)}{2E^{2}}
  \frac{C_{F}}{N_{c}}
   f_{\pi}f_{B}
   \int d\xi\phi_{B}(\xi)
   \int d u
\nonumber \\
&  & 
\times
 \left[
  \frac{
    \phi_{\pi}^{P}(u)\left[
      \eta+(1-2\eta)\xi-(1-2\eta)\eta\bar{u}
    \right]
   + \phi_{\pi}^{p}(u) r_{\chi}
    \left[
      (1-2\eta)\eta\bar{u}-(1-\eta)\xi
    \right]}
    {\eta^{2}\xi\bar{u}(\xi-\eta\bar{u})}
  \right]
\nonumber \\
 &  & 
+ \frac{\pi\alpha_{s}(t)}{2E^{2}}
  \frac{C_{F}}{N_{c}}
  f_{\pi}f_{B}
  \int d\xi\bar{\phi}_{B}(\xi)
  \int d u\phi_{\pi}^{P}(u)
  \left[
      \frac{\xi-(1-2\eta)\eta\bar{u}-\eta}
           {\eta\xi\bar{u}(\xi-\eta\bar{u})}
  \right]
\nonumber \\
 &  & 
+  \frac{\pi\alpha_{s}(t)}{2E^{2}}
   \frac{C_{F}}{N_{c}}
    f_{\pi}f_{B} 
    r_{\chi}^{\pi}
    \int d\xi\phi_{B}(\xi)
    \int d u_{q}\int d u_{\bar{q}}\int d u_{g}
     \delta(1-u_{q}-u_{\bar{q}}-u_{g})
\nonumber \\
&  & 
\times
   \frac{\phi_{\pi}^{3p}(u_{q},u_{\bar{q}},u_{g})}{u_{q}u_{\bar{q}}u_{g}}
    \left[
       \frac{\bar{\eta}\xi-(1-\eta(2-u_{g}))\eta u_{\bar{q}}}
             {\eta^{2}\xi\bar{u}_{g}(\xi-\eta u_{\bar{q}})}
     \right]
\,.
\nonumber 
\end{eqnarray}
The form factor $F_{+}^{B\pi}(q^{2})$ is analytic in the whole range of $q^2$, $0\leq q^2\leq q_{\max}$, 
$q^2_{\max}=26.42$ GeV$^2$ for $m_B=5.28$ GeV and $m_\pi=0.14$ GeV.
Under the maximal recoil limit, $\eta_{\max}=1$, 
the form factor $F_{+}^{B\pi}$ becomes
\begin{eqnarray}
\lim_{q^2\to 0}F_{+}^{B\pi}(q^2) 
& = & 
\frac{\pi\alpha_{s}(t)}{2E^{2}}
  \frac{C_{F}}{N_{c}}
   f_{\pi}f_{B}
   \int d\xi\phi_{B}(\xi)
   \int d u
   \left[
       \frac{
       \phi_{\pi}^{P}(u)\bar{\xi}+
       \phi_{\pi}^{p}(u)(1-r_{\chi}^{\pi})\bar{u}
            }
         {\xi\bar{u}(\xi-\bar{u})}
    \right]
\\
&  & 
+  \frac{\pi\alpha_{s}(t)}{2 E^{2}}
   \frac{C_{F}}{N_{c}}
    f_{\pi}f_{B}
    \int d\xi\bar{\phi}_{B}(\xi)
    \int du\phi_{\pi}^{P}(u)
    \left[
        \frac{\bar{u}-\bar{\xi}}{\eta\xi\bar{u}(\xi-\eta\bar{u})}
    \right]
\nonumber \\
 &  & 
+ \frac{\pi\alpha_{s}(t)}{2 E^{2}}
   \frac{C_{F}}{N_{c}}
   f_{\pi}f_{B}
   r_{\chi}^{\pi}
   \int d\xi\phi_{B}(\xi)\int du_{q}\int du_{\bar{q}}\int du_{g}
    \delta(1-u_{q}-u_{\bar{q}}-u_{g})
\nonumber \\
&  & 
 \times
  \frac{\phi_{\pi}^{3p}(u_{q},u_{\bar{q}},u_{g})}{u_{q}u_{g}}
  \left[
      \frac{1}{\xi(\xi-u_{\bar{q}})}
  \right]
\,.
\nonumber 
\end{eqnarray}
The dependence of strong coupling constant 
$\alpha_{s}(t)$ on $t\propto\sqrt{\eta}$
restrict the effective range of $q^{2}$ to be $0\leq q^{2}\leq16\,\text{GeV}^{2}$, 
in which the values of $\alpha_{s}$ are varying within $0.33\sim 0.48$.

\section{Numerical Calculations}
We apply previous results for numerical calculations of the form factor $F^{B\pi}_{+}(q^2)$ 
for $\bar{B}^0\to \pi^- l^+\bar{\nu}_l$ decays. 
The shape parameter $\omega_B$ of the $B$ meson DA and the value of $|V_{ub}|$ will be determined.
The $q^2$ shape of the form factor $F^{B\pi}_{+}(q^2)$ is given according to the value of $\omega_B$.
\subsection{analysis method} 
In order to explore the full range $q^2$ shape of $F_+(q^2)$ based on our calculations,
we employ a statistic analysis method.
We assume that the form factor $F_+(q^2)$ can smoothly vary with $q^2$ from small $q^2$ region to large $q^2$ region.
We first collect 
17 data points of $F_+(q^2)$ at $0\leq q^2\leq 16$ GeV$^2$  and $\Delta q^2=1$GeV$^2$.
We denote these 17 data points as the $C^f$ set of data points.  
Based on the $C^f$ set of data points, 
we derive a fit formula for $F_+(q^2)$ by a minimum $\chi^2$ fit
according to a $\chi^2_{FF}$ function.
The fit form factor $F^{\text{fit}}_+(q^2)$ is a function of $q^2$ with 9 parameters, $a_i$, $i=0,1,\cdots, 8$, as
\begin{eqnarray}
F^{\text{fit}}_+(x)=\sum_{i=0}^8 a_{i} x^{i}\,.
\end{eqnarray} 
The $\chi^2_{FF}$ function is defined as
\begin{eqnarray}
\chi^2_{FF}=\sum_i^N\frac{(F^{\text{fit}}_+(q^2_i)-F^{\text{data}}_+(q^2_i))^2}{\sigma^2_i}\;,
\end{eqnarray}
where $i$ denotes the $i$-th bin and $N$ is the total bin number of data points.
$q^2_i$ and $\sigma_i$ are the reference $q^2$ value 
and the uncertainty for the $i$-th bin of the used data set. 
$F^{\text{fit}}_+(q^2_i)$ is the value of $F^{fit}_+(x)$ at $x=q^2_i$.
$F^{\text{data}}_+(q^2_i)$ is the $i$-bin data. 
A $8\%$ error is taken for the input parameters
($\alpha_s$, $f_\pi$, $f_B$, $r_{\chi}$). 
See more detailed explanations about this $8\%$ error in next Subsection.
  
The correctness of the $q^2$ shape of $F_+^{\text{fit}}(q^2)$ for $q^2> 16$ GeV$^2$
is estimated by comparing 
the predictions of $F^{\text{fit}}_+(q^2)$ for $q^2>16$ GeV$^2$
to the LQCD calculations by the FNAL collaboration 
\cite{Arnesen:2005ez,Okamoto:2005zg, Bailey:2008wp}
and by the HPQCD collaboration \cite{Dalgic:2006dt}, respectively.


The fit form factors is used to make a prediction for the branching ratio 
denoted as $Br^{\text{th}}_{\text{SL}}$ for $\bar{B}\to\pi l\bar{\nu}_l$.
The $\omega_B$ and $|V_{ub}|$ are determined by comparing the predictions and experimental data, 
$Br^{\text{exp}}_{\text{SL}}$.
This is equivalent to a fit for the value of $|V_{ub}F_+(0)|$.
Because the $q^2$ shape of $F_+(q^2)$ is fixed by Eq.~(\ref{F+total}), theoretically,
the normalization point $|V_{ub}F_+(0)|$ can be determined by the fitting method.
The $|F_+(0)|$ and $|V_{ub}|$ are determined separately.
Our best fit to the central value of $Br^{\text{exp},full}_{\text{SL}}=(1.36\pm 0.09)\times 10^{-4}$ (PDG08) 
\cite{Amsler:2008zz} gives 
\begin{eqnarray}
\label{eq:result-full}
|V_{ub}F_{+}(0)| & = & (6.471\pm 0.196_{\text{exp}}\pm 0.523_{\text{th}})\times 10^{-4}
\nonumber\,,\\
 |F_+(0)|&=& {\displaystyle 0.164 {^{+0.010}_{-0.009}}|_{\alpha_s} {\pm 0.008}|_{f_B} }
\\
       &=& {\displaystyle 0.164 {{\pm 0.013}}}
\,,\nonumber\\
|V_{ub}|&=& (3.946{^{+ 0.128}_{-0.133}}_{\text{exp}}{^{+ 0.359}_{-0.282}}_{\text{th}})\times 10^{-3}
\,,\nonumber\\
\omega_B &=& 0.315 \text{GeV}
\,.\nonumber
\end{eqnarray}
The errors are experimental (exp) and theoretical (th).
The theoretical errors  are only considering those uncertainties from $\alpha_s$ and $f_B$. 
$|F_+(0)|$ is calculated according to the fit form factor and the fitted value $\omega_B$. 
$|V_{ub}|$ is determined by a combination of $|V_{ub}F_+(0)|$ and  $|F_+(0)|$. 
The central value of $|V_{ub}F_+(0)|$ is determined by a best fit to the central value 
of the $Br^{\text{exp}}_{\text{SL}}$.
The determinations of the fit form factor and corresponding $|V_{ub}|$ will be respectively described in detail latter. 

The $C^f$ form factor $F^{C^f}_+(q^2)$ is perturbatively meaningful for $q^2\leq 16$ GeV$^2$.
We perform an similar analysis as above by using the partial branching fraction 
$Br^{\text{exp}, q^2<16\text{GeV}^2}_{\text{SL}}=(0.93\pm 0.06)\times 10^{-4}$. 
\footnote{The value is quoted from the ICHEP08 averages given in the online update at
 http://www.slac.stanford.edu/xorg/hfag by the HFAG colaboration.
Note that the full branching fraction $Br^{\text{exp},\text{full}}_{\text{SL}}=(1.34\pm 0.08)\times 10^{-4}$ 
has been updated by HFAG at 2008}
The best fit result is given by 
\begin{eqnarray}
\label{eq:result-partial}
|V_{ub}F_{+}(0)| & = & (7.34\pm 0.233_{\text{exp}}\pm 0.597_{\text{th}})\times 10^{-4}
\,,\\
 |F_+(0)|&=& {\displaystyle 0.188 {^{+0.011}_{-0.010}}|_{\alpha_s} {\pm 0.009}|_{f_B} }
\nonumber\\
       &=& {\displaystyle 0.188 {{\pm 0.014}}_{\text{th}}}
\,,\nonumber\\
|V_{ub}|&=& (3.904{^{+ 0.124}_{-0.128}}_{\text{exp}}{^{+ 0.356}_{-0.279}}_{\text{th}})\times 10^{-3}
\,,\nonumber\\
\omega_B &=& 0.281 \text{GeV}
\,,\nonumber
\end{eqnarray}
The errors are the same as explained previous. 
We note that our best fit value $|V_{ub}F_+(0)|=6.5\times 10^{-4}$ 
is lower than the value \cite{Ball:2006jz},
which was obtained by a fitting to the branching fraction and $q^2$ spectrum data by BaBar collaboration 
\cite{Aubert:2006px},
\begin{eqnarray*}
|V_{ub}F_{+}^{B\pi}(0)| & = & (9.1\pm 0.3\pm 0.6)\times10^{-4}
\,,
\end{eqnarray*}
where the first error is from the uncertainty on $Br(\bar{B}\to\pi l\bar{\nu_{l}})$,
and the second from the parametrization of shape of the form factor
versus $q^{2}$. 
The $|V_{ub}F_{+}(0)|$ in Eqs.~(\ref{eq:result-full}) and (\ref{eq:result-partial}) is
consistent with the result $|V_{ub}F_{+}(0)|=(7.6\pm 1.9)\times 10^{-4}$ 
obtained with SCET and QCD factorization \cite{Bauer:2004tj}.

The analysis procedure is the following. 
We first choose a reference value of $\omega_B$ to prepare the $C^f$ set of data point for $F_+$.
Using this $C^f$ set of data, the fit form factor is determined according to 
the $\chi^2_{FF}$ function.
The fit form factor is then used to make a prediction for the branching fraction for $\bar{B}\to\pi l\bar{\nu}_l$
denoted as $Br^{fit}_{\text{SL}}$ by varying the value of $|V_{ub}|$.
A best fit for the predicted branching fraction 
and experimental one is performed according to the $\chi^2_{Br}$ function
\begin{eqnarray}
\chi^{2}_{Br}=\frac{(Br^{fit}_{\text{SL}}-Br^{exp}_{\text{SL}})^2}{\sigma^2}
\end{eqnarray}
where $\sigma$ is the experimental error calculated 
by adding the systematical and statistical errors in quadrature.
Once a best fit is found, $|V_{ub}|$ is determined.
Otherwise, another $\omega_B$ is chosen and the whole procedure is run from start, again.

\subsection{form factors}
The values of $\alpha_{s}(t)$ at different values of $q^{2}$ are calculated by the program \cite{Abe:2004zm}
and given in Table \ref{tab:strong_alpha}.
The program \cite{Abe:2004zm} has considered experimental and theoretical uncertainties. 
The scale $t$ is defined as the scale
of the internal virtual gluon,
which are assumed to carry a hard-collinear energy. 
$t=\sqrt{\eta\Lambda_{h}m_{B}}$ with
$\Lambda_{h}=0.5$ GeV and $m_{B}=5.28$ GeV.  

\begin{table}[t]

\caption{The values of $\alpha_{s}$ at $q^{2}$, 0$\leq q^{2}\leq 16\,\text{GeV}^{2}$, are calculated 
by the program \cite{Abe:2004zm}.}
\label{tab:strong_alpha}

\begin{tabular}{cccccccccc}
\hline 
$q^{2}(\text{GeV}^{2})$ & 0 & 2 & 4 & 6 & 8 & 10 & 12 & 14 & 16\tabularnewline
\hline
\hline 
$\alpha_{s}(t)$ & $0.333$ & $0.340$ & $0.348$ & $0.357$ & $0.368$ & $0.400$ & $0.421$ & $0.447$ & $0.484$\tabularnewline
$\delta\alpha_{s}(t)$ & $^{+0.020}_{-0.018}$ & $^{+0.20}_{-0.019}$ & $^{+0.022}_{-0.019}$ 
& $^{+0.023}_{-0.021}$ & $^{+0.025}_{-0.022}$ & $^{+0.032}_{-0.028}$ & $^{+0.036}_{-0.031}$ 
& $^{+0.043}_{-0.036}$ & $^{+0.053}_{-0.043}$\tabularnewline
\hline
\end{tabular}
\end{table}

The uncertainties on the $F_+$ could be 
(1) phenomenological models for meson DAs, 
(2) input parameters ($\alpha_{s}$, $f_{\pi}$, $f_{B}$, $r_{\chi}^{\pi}$, $\omega_{B}$, $\eta_3$, and $\omega$). 
The uncertainty on phenomenological models is fixed
by taking models determined from other processes. 
The twist-2 and twist-3 meson DAs takes the following models  
\begin{eqnarray*}
\phi_{\pi}^{P}(u) 
& = & 
\phi_{\pi}^{p}(u)=\phi_{\pi}^{\sigma}(u)=6u(1-u)
\,,
\\
\phi_{\pi}^{3p}(u_{q},u_{\bar{q}},u_{g}) 
& = & 
   360\eta_{3}u_{q}u_{\bar{q}}u_{g}^{2}(1+\frac{\omega}{2}(7u_{g}-3))
\,,
\end{eqnarray*}
where $\eta_{3}=0.015$ and $\omega=-3.0$ are used. 
The chiral factor is set
$r_{\chi}^{\pi}=0.76\,\text{GeV}$. 
The total uncertainties on the input parameters are assumed constant of $8\%$.
The recent LQCD calculations for $f_B$ are
$f_B=204^{+37}_{-28}$ MeV (CP-PACS) \cite{AliKhan:2001jg},
$f_B=190\pm 14$ MeV (JL-QCD) \cite{Yamada:2001xp},
$f_B=206\pm 10$ MeV (ALPHA) \cite{Heitger:2003ga},
$f_B=216\pm 11$ MeV (HPQCD) \cite{Gray:2005ad,Shigemitsu:2005wv},
$f_B=214\pm 9$ MeV (Guo-Weng) \cite{Guo:2006nt}.
Except of the CP-PACS result with $16\%$ errors, 
the other calculations contain about $5\%$ uncertainties. 
In our calculations, we choose $f_B=200$ MeV as a reference value
and associate a $5\%$ error on it.
The chiral enhanced contributions are about $20\%$ in our calculations.
We choose $r_\chi^{\pi}=2m_\pi^2/(m_B(m_u+m_d))=0.76$ GeV. 
This corresponds to use $(m_u+m_d)=9.8$ MeV \cite{Amsler:2008zz},  
$m_\pi=140$ MeV, and $m_B=5.28$ GeV.
The pion decay constant $f_{\pi^\pm}=(130.4\pm 0.04 \pm 0.02)$ MeV and 
$f_{\pi^0}=(130\pm 5)$ MeV \cite{Amsler:2008zz}. 
The errors on $f_{\pi^-}$ can be neglected.
The uncertainties on $\alpha_s$ are $5-10\%$. 
There are about $10\%$ errors associated with $r_{\chi}$.
Because the chiral contributions with $r_{\chi}$ are about $20\%$ of the leading twist (twist-2) contributions,
the errors on $r_{\chi}$ are about $0.2\times 0.1\simeq 2\%$ and can be neglected safely.
The constant $8\%$ errors on $F_+(q^2)$ is assumed according to the errors on $\alpha_s$ and $f_B$.
Instead of using $\bar{\Lambda}=m_B-m_b$ and $\lambda_B$ for the value of $\omega_{B}$,
we take $\omega_B$ as a free parameter
and propose to determine $\omega_B$ and $|V_{ub}|$ by the world branching fraction $Br^{exp}_{\text{SL}}$
for $\bar{B}\to \pi l\bar{\nu}_l$
\begin{eqnarray}
Br^{exp}_{\text{SL}}(\bar{B}^{0}\to\pi^{-}l^{+}\nu_{l})
=\left(
1.36\pm 0.06_{\text{stat}}\pm 0.07_{\text{sys}}
\right)
\times
10^{-4}
\end{eqnarray}
for $0\leq q^{2}\leq q_{max}^{2}$, 
where $q_{max}^{2}=26.42\,\text{GeV}^{2}$. 

17 bins of $C^f$ data points of $F_{+}^{B\pi}(q^{2})$ with 
$\Delta q^{2}=1\text{GeV}^{2}$ are prepared by taking a test value of $\omega_B$.
In our present case, we choose $\omega_B=0.46$ GeV as the first value.
The $F^{fit}_+(q^2)$ is then used to make a $\chi^2$ fit to these 17 bins of data.
The best fit is determined by requiring the $\chi^2_{FF}$ function to be minimal.
The fit form factor is then substituted in Eq.(\ref{eq:decay ray}) 
to make a prediction for the branching fraction, $Br^{th}_{\text{SL}}$ 
by varying the value of $|V_{ub}|$ in $2.8\times 10^{-3} \leq |V_{ub}|\leq 4.5\times 10^{-3}$.
The best fit is found by using the $\chi^2_{\text{Br}}$ function to calculate the minimal $\chi^2$ value. 
Our best fit gives 
\begin{eqnarray}
\omega_B=0.315\pm 0.000\,\text{GeV}
\;,\;
|V_{ub}|=3.946^{+0.006}_{-0.002}\times 10^{-3}
\end{eqnarray}
under $\chi^2_{\text{Br}}\leq 0.001$.
The errors on the fit value of $\omega_B$ and $|V_{ub}|$ are statistic.
The fit form factor is assumed to have 9 parameters
\begin{eqnarray}
\label{eq:fit-FF+}
F_{+}^{\text{fit}}(x) & = & \sum_{i=0}^{i=8}a_i x^i
\,.
\end{eqnarray}
The fitting of $F_{+}^{\text{fit}}(x)$ to the 17 data points is performed 
by using different numbers of parameters, $1$ to $9$.
The minimal $\chi^2$ value of the $\chi^2_{\text{FF}}$ function is required to derive a best fit.
The $\chi^2_{\text{FF}}/N$ is equal to $0.05/(17-5)$ by assuming an $8\%$ constant errors on the $C^f$ data set.
The values of parameters $a_i$, $i=0 \cdots 8$, are given in Table \ref{tab:fit form factors}.
The errors on $a_i$, $i=0 \cdots 8$, are statistic due to the best fit of $\omega_B$.
We denote the fit form factor as $F^{C^f}_+(q^2)$.
\begin{table}[t]
\caption{The values of parameters of $F_{+}^{fit}(q^{2})$ and the best fit value of $\omega_B$ in unit of GeV.
The best fit means $\chi^2_{\text{Br}}\leq 0.001$.
The errors on $\omega_B$ and $a_i$, $i=0,\cdots,8$, are statistic.
}
\label{tab:fit form factors}
\begin{tabular}{|c|ccccccccccc|}
\hline 
method & $\frac{\chi^2_{\text{FF}}}{N}$&$\omega_B\times10^{1}$ & $a_0\times10^{1}$ & $a_1\times10^{3}$ & $a_2\times10^{3}$ 
& $a_3\times10^{4}$ & $a_4\times10^{5}$ & $a_5$ & $a_6$ & $a_7$ & $a_8$\tabularnewline
\hline
\hline 
full        & $\frac{0.05}{12}$ & $3.151^{+3}_{-3}$ & $1.636^{+2}_{-2}$ & $8.403^{+12}_{-10}$ 
                   & $1.903^{+2}_{-2}$ & $-1.873^{+2}_{-2}$ & $1.407^{+2}_{-2}$ & 0 & 0 & 0 & 0\tabularnewline
\hline
partial $q^2\leq 16$ GeV$^2$       & $\frac{0.0}{12}$ & $2.810^{+5}_{-5}$ & $1.881^{+4}_{-4}$ & $9.773^{+23}_{-22}$ 
                   & $2.185^{+5}_{-4}$ & $-2.143^{+5}_{-4}$ & $1.625^{+3}_{-4}$ & 0 & 0 & 0 & 0\tabularnewline
\hline
\end{tabular}
\end{table}

\begin{figure}[t]
\includegraphics[scale=0.8]{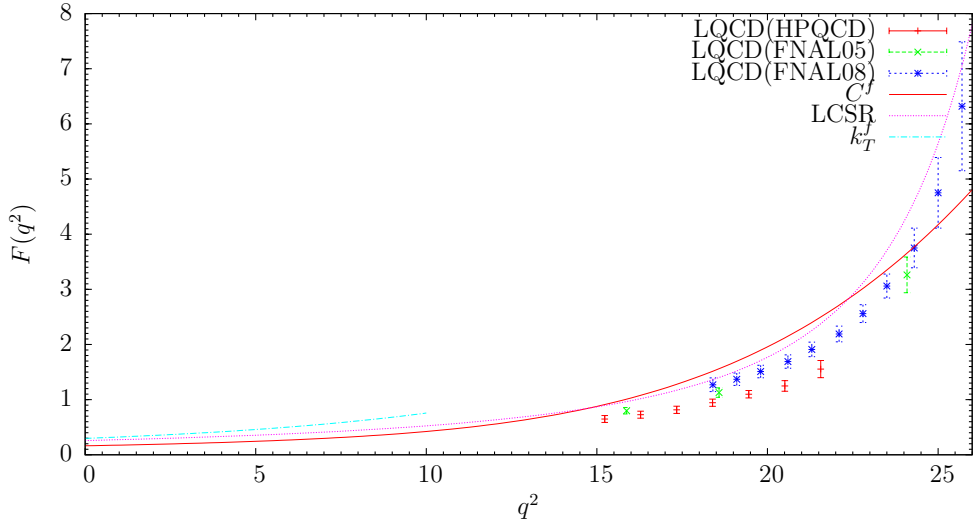}
\caption{Comparisons between $C^f$(solid line), $k^f_T$(dash-dot-dash line), LCSR(dot line), 
, LQCD-FNAL (data point), and LQCD-HPQCD (data point) form factors .}
\label{fig:FFsComparison}
\end{figure}

By $F_{+}^{Cf}(q^{2})$, we compare our calculations
with the same form factor calculated by the other approaches,
including the $k_{T}^{f}$ method \cite{Kurimoto:2001zj}, 
the LCSR method \cite{Ball:2001fp},
the LQCD-HPQCD method \cite{Dalgic:2006dt} 
the LQCD-Fermilab/MILC/2005 (FNAL05) method \cite{Arnesen:2005ez,Okamoto:2005zg}
and the LQCD-Fermilab/MILC/2008 (FNAL08) method \cite{Bailey:2008wp}. 
The comparisons are shown in Fig. \ref{fig:FFsComparison}. 
The LCSR form factor \cite{Ball:2001fp} is expressed as
\begin{eqnarray}
F^{\text{LCSR}}_+(q^2)=\frac{r_1}{1-\frac{q^2}{m_1^2}}+\frac{r_2}{1-\frac{q^2}{m_{\text{fit}}^2}}
\,,
\end{eqnarray}
where $r_1=0.744$, $r_2=-0.486$, $m_1=5.32$ GeV, $m^2_{\text{fit}}=40.73$ GeV$^2$.
Both LCSR and $C^f$ form factors are larger than the LQCD form factors at most range of $q^2$.
The $k^f_T$ form factor is larger than the $C^f$ and LCSR form factors in the available range, $q^2\leq 10$ GeV$^2$.
We find that the $\chi^2/N$ values of $F^{C^f}_+(q^2)$ 
versus the FNAL(=FNAL05+FNAL08) and HPQCD data are $54.8/15$ and $271.1/7$, respectively. 
It is seen that $F_+^{C^f}(q^2)$ for $q^2\geq 16$ GeV$^2$ is close to the FNAL data 
but has a large deviation with the HPQCD data.

The twist-2, $F_{+}^{B\pi,tw2}$, and twist-3, $F_{+}^{B\pi,tw3}$,
components of the form factor $F_{+}^{B\pi}$ are calculated at different
$q^{2}$,
$0\leq q^{2}\leq16\,\text{GeV}^{2}$ in Table. \ref{tab:form_factors}.
It is seen that $F_{+}^{B\pi,tw2}>F_{+}^{B\pi,tw3}
(=F_{+}^{B\pi,tw3,ps}+F_{+}^{B\pi,tw3,pt}+F_{+}^{B\pi,tw3,3p}+F_{+}^{B\pi,tw4,\bar{B}})$
in the considered range of $q^{2}$. 
The power expansion is meaningful.
We note that $F_+^{C^f}(0)=0.164$ is obtained.
This is close to the finding by the SCET analysis for $B\to\pi\pi$ decays \cite{Bauer:2004tj},
that $F_+^{\text{SCET}}(0)=0.170$.

\begin{table}[t]

\caption{The values of different contributions of $|F_{+}^{B\pi}(q^{2})|$
at $q^{2}$, 0$\leq q^{2}\leq16\,\text{GeV}^{2}$. $\omega_{B}=0.315\,\text{GeV}$
is used.}
\label{tab:form_factors}
\begin{tabular}{|c|ccccccccc|}
\hline 
$q^{2}$(GeV$^{2}$) & 0 & 2 & 4 & 6 & 8 & 10 & 12 & 14 & 16\tabularnewline
\hline
\hline 
$|F_{+}^{B\pi,tw2}(q^{2})|\times 10^{3}$       & 175 & 205 & 240 & 286 & 351 & 444 & 582 & 787 & 1100\tabularnewline
$|F_{+}^{B\pi,\bar{B}}(q^{2})|\times 10^{4}$   & 129 & 152 & 180 & 218 & 270 & 347 & 463 & 640 & 919\tabularnewline
$|F_{+}^{B\pi,tw3,ps+pt}(q^{2})|\times 10^{4}$ & 196 & 248 & 318 & 417 & 568 & 810 & 1213 & 1911 & 3202\tabularnewline
$|F_{+}^{B\pi,tw3,3p}(q^{2})|\times 10^{4}$    & 106 & 127 & 154 & 193 & 255 & 360 & 547 & 900 & 1615\tabularnewline
total$\times 10^{3}$                           & 162 & 188 & 219 & 259 & 314 & 393 & 508 & 679 & 943 \tabularnewline
\hline
\end{tabular}
\end{table}

We note that $F^{C^f}_+(0)=0.164$ is smaller than
$F_+(0)=0.25\sim 0.30$ derived by other theories (such as, LCSR, LQCD, $k^f_T$,), and experiments. 
As was firstly found by Bauer {\it et al}\cite{Bauer:2004tj},
there is an inconsistency  between the small $F^{\text{SCET},Cf}_+(0)=0.17$ implied in $B\to\pi\pi$ decays
and the large $F_+(0)=0.25\sim 0.30$ derived by other theories and experiment. 
For example, $F^{B\pi}_+(0)=0.28$ was used in the QCD factorization for $B\to\pi\pi$ decays
and the QCD factorization predictions for the branching ratios of $B\to\pi\pi$ decays
were inconsistent with the experimental data.
On the other hand, if $F^{\text{SCET}}_+(0)=0.17$, 
then the QCD factorization based on SCET can explain the experimental data consistently.
According to our calculations, 
the small $F^{Cf}_+(0)=0.164\pm 0.013$ can also explain 
the semi-leptonic decay $\bar{B}\to\pi l\bar{\nu}_l$ 
with a value of $|V_{ub}|$ consistent with the world averaged value.
Recently, a general parameterization approach 
\cite{Becirevic:1999kt,Fukunaga:2004zz,Arnesen:2005ez,Flynn:2007qd,Flynn:2007ii} 
has been used to study the $q^2$ shape of $F_+(q^2)$.
Among of different parameterization approaches 
\cite{Becirevic:1999kt,Fukunaga:2004zz,Arnesen:2005ez,Flynn:2007qd,Flynn:2007ii},
Arneson et al \cite{Arnesen:2005ez} found that 
it is possible to consistently explain the branching ratio and BaBar differential rate data 
by using the SCET value $F^{\text{SCET}}_+(0)=0.17$.  
The other parameterization approaches \cite{Fukunaga:2004zz,Flynn:2007qd,Flynn:2007ii} 
employed the LCSR value $F_+(0)=0.26$ as an input.
Whether the value of $F_+(0)$ is small or large is a controversial topic,
it needs more theoretical and experimental works to clarify this problem.

The fit result
$\omega_{B}=0.315\text{GeV}$
implies $\bar{\Lambda}=0.471$ GeV or $\lambda_{B}=0.315\text{GeV}$. 
We note that $\lambda_{B}=0.315\text{GeV}$ is lower than $0.460\pm 0.110$ GeV by Braun {\it et al}\cite{Braun2004},
$0.454$ GeV by Grozin and Neubert \cite{Grozin:1996pq},
$0.479\pm 0.089$ GeV by Lee and Neubert \cite{Lee2005}, 
$0.600$ GeV by Ball \cite{Ball2003}.
$m_b=m_B-\bar{\Lambda}=(5.28-0.471)\,\text{GeV}=4.57\,\text{GeV}$,
is consistent with the $m_b$ determined by different schemes,
$m_{b}^{1S}=4.701\pm 0.030$ GeV (1S scheme)\cite{Barberio2007},
$m_{b}^{SF}=4.630\pm 0.060$ GeV (shape function scheme) \cite{Barberio2007}, 
and $m_{b}^{kin}=4.613\pm 0.035$ GeV (kinetic scheme)\cite{Barberio2007}, 
but is larger than 
$m_{b}^{\bar{MS}}=4.164\pm 0.025$ GeV (MS bar scheme)\cite{Kuhn2007}. 
At the chiral point, 
$E_{\pi} \simeq m_{\pi}$ and 
$q^{2}(E_{\pi})=26.42$ GeV$^2$,
the chiral perturbation theory predicts that
\begin{eqnarray}
F^{\text{Chiral}}_+(q^2)=\frac{g f_B m_B}{2 f_\pi (E_\pi + m_{B^*}-m_B)}
\,,
\end{eqnarray}
where $g$ is the $B^{*}B\pi$ coupling.
The fit form factor $F^{Cf}_+(q^2)$ predicts $F^{Cf}_+(26.42)=5.12$,
which implies $g=0.234\pm 0.019$ with a theoretical error
by using $m_{B^*}-m_B=45.75\pm 0.35$ MeV.
The extracted $g=0.234\pm 0.019$ is consistent with $g=0.22\pm 0.07$ determined by the FNAL collaboration
\cite{Bailey:2008wp} and $0< g <0.45$ by the HPQCD collaboration  \cite{Dalgic:2006dt},
but is lower than the usually employed value $g=0.51$ proposed by Stewart{\it et al }\cite{Arnesen:2005ez}.

\subsection{determination of $V_{ub}$}

The differential decay rate for 
$\bar{B}^{0}\to\pi^{-}l^{+}\bar{\nu_{l}}$,
under the approximation that the lepton masses are vanishing, is
given by 
\begin{eqnarray}
\label{eq:decay ray}
\left(\frac{d\Gamma}{dq^{2}}\right)_{\text{th}} 
& = & 
\frac{G_{F}^{2}|V_{ub}|^{2}}{24\pi^{3}}|p_{\pi}|^{3}|F_{+}^{B\pi}(q^{2})|^{2}
\end{eqnarray}
where $p_{\pi}$ is the momentum of pion in the $B$ meson rest frame.
The branching ratio 
$Br(B^{0}\to\pi^{-}l^{+}\bar{\nu_{l}})$ 
is expressed in terms of $(d\Gamma/d\eta)_{\text{th}}$
\begin{eqnarray}
Br(B^{0}\to\pi^{-}l^{+}\bar{\nu_{l}}) 
& = & 
\tau_{B^{0}}
\int_{0}^{q_{max}^{2}}dq^{2}
 \left(\frac{d\Gamma}{dq^{2}}\right)_{\text{th}}
\,.
\label{eq:branchingratio}
\end{eqnarray}
where 
$\tau_{B^{0}}=(1.530\pm 0.009)$ps 
is the life time of $\bar{B}^{0}$ meson \cite{Amsler:2008zz}. 

To determine $|V_{ub}|$, 
we employ 
\begin{equation}
R(q_{max}^{2})
\equiv
|V_{ub}|^{-2}
\int_{0}^{q_{max}^{2}}dq^{2}
\left(\frac{d\Gamma}{dq^{2}}\right)_{\text{th}}
\,.
\end{equation}
The result is 
\begin{eqnarray}
R(q_{max}^{2}) 
& = & 
(5.709\pm 0.913_{\text{th}})ps^{-1}\,.
\end{eqnarray}
By substituting 
$R(q_{max}^{2})$ 
into Eq. (\ref{eq:branchingratio}),
we can determine 
\begin{eqnarray}
\label{eq:Vub}
|V^{C^f}_{ub}| & = & (3.946\pm 0.117_{\text{exp}}\pm 0.312_{\text{th}})\times10^{-3}\,.
\end{eqnarray}
The first error on $|V_{ub}|$ is from branching ratio
and the second error is from $F_{+}(0)$. 
The theoretical uncertainty is $8\%$ in the same level with that of the inclusive method. 
For comparison, the theoretical errors are $15\%$ in the LCSR method, 
$10-14\%$ in the LQCD \cite{Amsler:2008zz},
and $12\%$ in the parameterization approach (PA) 
\cite{Becirevic:1999kt,Fukunaga:2004zz,Flynn:2007qd,Flynn:2007ii}. 
The $|V_{ub}|$ in Eq. (\ref{eq:Vub}) is consistent with 
the inclusive value
$|V^{\text{incl}}_{ub}|=(4.12\pm0.15_{\text{exp}}\pm0.40_{\text{th}})\times10^{-3}$ 
from 
$B\to X_{u}l\bar{\nu_{l}}$
\cite{Amsler:2008zz} 
and the world averaged $|V^{\text{avg}}_{ub}|=(3.95\pm 0.35)\times 10^{-3}$ \cite{Amsler:2008zz}.
We note that the world averaged exclusive value is $|V^{\text{excl}}_{ub}|=(3.5\pm 0.6)\times 10^{-3}$ \cite{Amsler:2008zz}.
The discrepancy between the values of $|V_{ub}|$ determined from exclusive and inclusive methods
has raised a lot of discussions in literature \cite{Ball:2006jz}.
However, we note that a smaller inclusive value $|V^{\text{incl08}}_{ub}|=(3.98\pm 0.15 \pm 0.30)\times 10^{-3}$
with experimental and theoretical errors  
was obtained by Neubert \cite{Neubert2008}.
This shows that there may not exist such a discrepancy.
If we only employ the partial branching ratio for $q^2< 16$ GeV$^2$, 
a smaller $|V_{ub}|$ will be obtained as given in Eq.~(\ref{eq:result-partial}). 
The analysis method is similar to the above and skipped here.  


\subsection{$q^2$ shape}
\begin{figure}[t]

\includegraphics[scale=0.7]{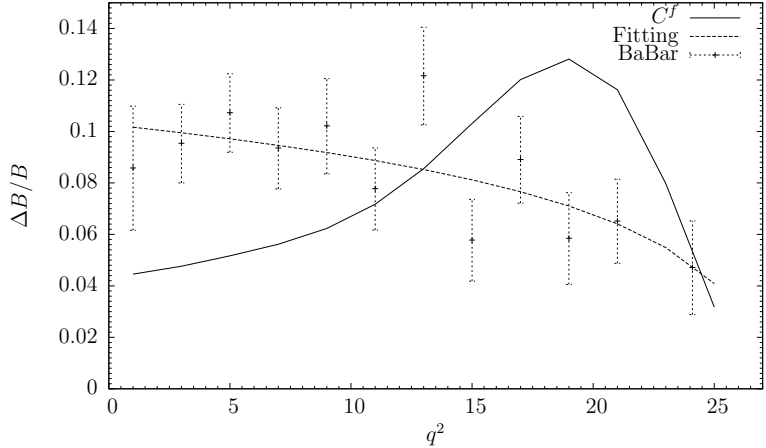}
\caption{$q^2$ shape comparisons between $C^f$ form factor (the solid line) and experimental data. 
The dash line is a fitting to the experimental data.}
\label{fig:q2spect}
\end{figure}

BaBar has measured the differential 
rate for $\bar{B}\to\pi l\bar{\nu_{l}}$ versus $q^{2}$ with good accuracy \cite{Aubert:2005cd, Aubert:2006px}.
The $q^2$ shape of the differential rate relates to the $q^2$ shape of $F^{B\pi}_+(q^2)$. 
It is interesting to compare our prediction for the spectrum with the data. 
Based on $F_{+}^{C^f}$, 
the $q^{2}$ spectrum $\Delta Br/Br$ is calculated as shown in 
Fig. \ref{fig:q2spect}.
We observe that the predicted shape of
$q^{2}$ spectrum ($C^f$ $q^2$ shape) is inconsistent with the experimental $q^2$ spectrum.
The $\chi^2$ is $86.6$ for 12 degrees of freedom by assuming 
that the 12 bins of experimental data are completely uncorrelated.
Parameterization methods \cite{Becirevic:1999kt,Fukunaga:2004zz,Flynn:2007qd,Flynn:2007ii} 
have been widely used to determine the $q^2$ shape of $F_+(q^2)$
according to the $q^2$ shape data and theoretical inputs from LCSR and LQCD.
It is an important task to compare $F^{C^f}_+(q^2)$ with these parameterization form factors to investigate
their differences .
This is left to other places.
In summary, our calculations for the $q^2$ spectrum of the differential rate 
for $\bar{B}\to\pi l\bar{\nu}_l$ decays
can not accommodate with the $q^2$-spectrum of $\bar{B}\to\pi l\bar{\nu_{l}}$.
The difference can be understood by the scaling behavior of $F^{C^f}(q^2)$
and that of $(d\Gamma/dq^2)_{\text{th}}$,
$F^{C^f}_+(q^2)\propto \eta^{-(1+0.1 x)}$ and $(\Gamma^{-1}d\Gamma/dq^2)_{\text{th}}\propto \eta^{(1-0.2x)}$ 
for $0\leq q^2\leq 10 $ GeV$^2$ and 
$x=q^2/(1 GeV^2)$, 
and
$F^{C^f}_+(q^2)\propto \eta^{-2}$ and $(\Gamma^{-1}d\Gamma/dq^2)_{\text{th}}\propto \eta^{-1}$ for $q^2\geq 10$ GeV$^2$.
We note that the form factor implied in BaBar $q^2$ data can have a scaling 
$F^{\text{PA}}_+(q^2)\propto \eta^{-1.5}(x+\eta^{y})^{1/2}$ with $x=-0.897$ and $y=0.0281$.
This gives $\Gamma^{-1}d\Gamma/dq^2\propto x+\eta^{y}$.
The residual scaling factor $\eta^{y}$ in  $\Gamma^{-1}d\Gamma/dq^2$ has small effects for large $\eta$ (small $q^2$) 
and is important for small $\eta$ (large $q^2$). 

\section{Comparisons with other approaches}
Many progresses in theories for $B\to\pi$ form factors were obtained
in past years. Partial $O(1/m_{B})$ corrections have been calculated
in the $k_{T}$ factorization approach \cite{Kurimoto:2001zj,Huang:2004hw}
and LCSR \cite{Belyaev:1993wp, Khodjamirian:1998ji,Duplancic:2008ix}.
The unquenched quark effects  
\cite{Bowler:1999xn, Okamoto:2004xg,Shigemitsu:2004ft,Dalgic:2006dt,Bailey:2008wp}
were included in the lattice QCD method. 
An all order proof of the factorization of the form factors was given for leading
twist and twist-3 two parton contributions in the collinear factorization \cite{Nagashima:2002iw}
and in the soft collinear effective theory (SCET) \cite{Bauer2003}, respectively. 
Different strategies for a parametrization of form factors have been developed
\cite{Becirevic:1999kt,Fukunaga:2004zz,Arnesen:2005ez,Ball:2006jz,Flynn:2007qd,Flynn:2007ii, Ball:2007sw}.
We compare our result with various theoretical methods.

\subsection{Comparisons with $k_T^f$}
It is the end-point divergent problem that the $B\to\pi$ form factors
contain logarithmic end-point divergences at twist-2 and linear end-point
divergences at twist-3. 
To solve the divergent problem, 
the general
wisdom is to use $k_{T}$-regularization for logarithmic end-point divergences.
For linear end-point divergences, 
a threshold re-summation is needed, too. 
Twist-3 contributions dominate over the twist-2 ones in the $k_{T}$ factorization approach. 
This is known as a power expansion problem in $k_T^f$.
One possible solution to this problem is to employ twist-3 wave
functions with better end-point behaviors. 
However, this would introduce additional uncertainties. 

For comparison,
let's compare the values of $F_{+}^{B\pi,tw2}(0)$ calculated under
the $k_{T}^{R}$ and $\xi^{R}$. 
Under $k_{T}^{R}$, 
the result is
$F_{+,k_{T}^{R}}^{B\pi,tw2}(0)=0.120$ \cite{Kurimoto:2001zj}, 
where 
$f_{\pi}=0.13\,\text{GeV}$, $f_{B}=0.19\,\text{GeV}$, 
and $\omega_B=0.40$ GeV
are used. 
Under $\xi^{R}$, 
$F_{+,\xi^{R}}^{B\pi,tw2}(0)=0.124$ by using the same input parameters.
This shows that $\xi^{R}$ is as effective as $k_{T}^{R}$ at leading twist order.

The large value $F_+(0)$ is related to the power expansion method used in $k_T^f$.
In $k_T^f$, only $O(1/m_B)$ contributions from two parton Fock state of the pion were calculated.
In \cite{Kurimoto:2001zj}, $\phi^{\pi}_{p}(u)=1$ for the PS DA 
and $\phi_{\sigma}(u)=6u(1-u)$ for the PT DA of the pion were used.
The collinear expansion scheme proposed by Beneke and Neubert (the BN scheme) \cite{Beneke:2003zv} was employed. 
At $q^2=0$, the leading twist contributions lead to $F^{\text{tw2}}_+(0)=0.120$ 
and the subleading twist contributions result in $F^{\text{tw3}}_+(0)=0.177$.
It is seen that $F^{\text{tw3}}_+(0)/F^{\text{tw2}}_+(0)=1.5$.
In comparison, the $\xi^R$ calculations give $F^{\text{tw3}}_{+,\xi^R}(0)/F^{\text{tw2}}_{+,\xi^R}(0)=0.18$.
This shows that the twist-3 contributions in $k_T^f$ dominate over the twist-2 contributions.
On the other hand, the twist-3 contributions in $C^f$ are power suppressed as expected.
The authors in \cite{Kurimoto:2001zj} proposed a new counting rule for the linear divergence associated with the twist-3 terms
to explain this phenomenology. 
However, if higher order power corrections contain similar or even worse end point divergences, 
the power expansion would breaks down.
This is the power expansion problem in $k_T^f$ \cite{Huang:2004hw}.
Hwang et al \cite{Huang:2004hw} pointed out that 
intrinsic parton transverse momenta are effective for the power expansion problem.
The contributions related to the intrinsic transverse momenta are unknown, in principle. 
They proposed to use the model
\begin{eqnarray}
\Psi(x,\vec{b})=\int_{|k|<1/b}d^2 { \vec{k}}_{T}\exp(-i {\vec{k}}_T\cdot{\vec{b}})\Psi(x,{\vec{k}}_T)
\end{eqnarray}
for all the wave functions (including twist-2 and twist-4 $B$ meson WFs, and twist-2 and twist-3 two parton pion WFs) 
used in calculations.
$\vec{b}$ are the conjugate coordinates to the intrinsic transverse momenta $\vec{k}_T$.
Because the $\Psi(x,\vec{b})$ provides a better end-point suppression than the constant model for the twist-3 pseudoscalar pion WF,
the twist-3 contributions become smaller than the leading twist ones,
$F^{\text{tw3}}_+(0)/F^{\text{tw2}}_+(0)=1.5\to 0.7$.
However, it is still unclear how higher order power contributions,
such as twist-3 three parton and twist-4 contributions, can be included in the $k_T^f$ approach.
In addition, the introduction of intrinsic transverse momentum would arise a double counting problem
when higher Fock state contributions are considered, 
such as three parton Fock state contributions, or four parton Fock state contributions.
This is also unclear in $k_T^f$.
We argue that the power expansion problem in $k_T^f$ needs more efforts to clarify.
Before this problem has been solved, 
the large value $F^{k_T^f}_+(0)=0.297$ in $k_T^f$ could be over-estimated.
   
\subsection{Comparisons with SCET}
In SCET, the form factor $F_{+}^{B\pi}$ contain a factorizable and a nonfactorizable parts at leading
twist order, where the factorizable part is expressed in terms of a nonperturbative form factor, $\zeta^{B\pi}_+$ \cite{Bauer2003}.
By a fitting to the experimental data for the branching ratios of $B\to\pi\pi$ decays,
$F_{+}^{B\pi,\text{SCET}}(0)=0.170$ was found \cite{Bauer:2004tj}.
It is noted that our predicted value $F^{B\pi}_+(0)=0.164$ is close to the SCET result.
This insures that the form factor $F^{B\pi}_+(0)$ may not be as large as found from other approaches.
The values derived by LSCR and LQCD and $k_T^f$ are about $0.26-0.30$.
The consistency between $F_+^{C^f}(0)$ and $F_+^{\text{SCET}}(0)$ may not be an accident.
In the energetic limit, $E\gg m_\pi$, the $q(0)\bar{q}(z)$ in the matrix element 
$\langle 0|q(0)\bar{q}(z)|\pi\rangle$ for defining the DA,
becomes $q(0)\bar{q}(\lambda n/E)+\cdots$,
where $q(\lambda n/E)$ is similar to the collinear quark field $q_c(\lambda n/E)$ defined in the SCET$_I$.
The collinear factorization based on full QCD with energetic limit for light meson is likely equivalent to
the SCET$_I$ \cite{Nagashima:2002iw}. 
However, it needs further works to show the equivalence between $C^f$ and SCET$_I$ for $F_+^{B\pi}(q^2)$. 

The end-point divergences are also found in the SCET approach.
The zero bin subtraction regularization method was proposed for dealing with soft and end-point divergences
\cite{Manohar2007}.
Since the related physics about the subtracted quantities have not been given in the SCET language,
there still remained some uncertainties in this method. 

\subsection{Comparisons with LCSR}
The LCSR has been widely used for calculations of the form factors 
\cite{Chernyak:1977as,Chernyak:1990ag,Belyaev:1993wp,Khodjamirian:1997ub, Ball:1998tj, Khodjamirian:1998ji,Ball:2004ye,
Ball:2005tb, Khodjamirian:2006st, Duplancic:2008ix}. 
The LCSR method calculates $f_BF_+^{B\pi}(q^2)$ 
by matching the hadronic part to the partonic part
of the correlator under parton-hadron duality.
%
Let's first compare the LCSR form factor $F^{\text{LCSR}}_+(0)$ 
and the $C^f$ form factor $F^{C^f}_+(0)$ at twist-2 order.
Numerically, $|F^{\text{LCSR},tw2}_+(0)|=0.164\sim 0.170 $ \cite{Belyaev:1993wp, Khodjamirian:1998ji,Duplancic:2008ix}
and $|F^{C^f,tw2}_+(0)|=0.175$ ($\omega_B=0.315$ GeV) 
show that the LCSR and the $C^f$ calculations are consistent at twist-2 order.
At twist-3 level, we find that the ratio of twist-3 contributions to the twist-2 contributions
is equal to $|F^{\text{LCSR},tw3}_+(0)|/|F^{\text{LCSR},tw2}_+(0)|\simeq 80\%-100\%$ 
\cite{Khodjamirian:1998ji,Duplancic:2008ix}.
Similarly, the ratio  
in the $C^f$ approach gives $|F^{C^f,tw3}_+(0)|/|F^{C^f,tw2}_+(0)|\simeq 17\%$ ($\omega_B=0.315$ GeV).
We observe that twist-3 contributions in the LCSR form factor are more significant than those in the $C^f$ form factor.
This also explain why the difference between
 $|F_+^{C^f}(0)|=|F^{C^f,tw2}_+(0)+ F^{C^f,tw3}_+(0)|=0.164$ ($\omega_B=0.315$ GeV)
and $|F_+^{\text{LCSR}}(0)|=|F_+^{\text{LCSR},tw2}(0)+F_+^{\text{LCSR},tw3}(0)|=0.27$ (at $O(\alpha_s)$)
is so large.
Note that there are partial cancellations in the sum of the twist-2 and twist-3 contributions of the form factors.
It was also noticed that the soft contributions from end point region in the LCSR analysis are important 
\cite{Khodjamirian:1997ub,Kurimoto:2001zj}.
Because the twist-3 contributions have a more sensitive dependence on the end point behaviors of the pion DAs
than the twist-2 ones, 
this may explain why the twist-3 contributions are so significant in the LCSR form factor.
In LCSR calculation, the constant model for the PS DA of pion, $\phi_p(u)=1$, is used.
It is expected \cite{Khodjamirian:1997ub,Kurimoto:2001zj} that 
if the end point contributions can be properly dealt with by appropriate method, such as Sudakov factors,
the soft contributions can be reduced. 
We argue that the twist-3 contributions in the LCSR form factor could be overestimated.
A better power expansion method may solve this problem.

\subsection{Comparisons with BN collinear expansion}
An expansion scheme developed by Beneke and Neubert (BN) \cite{Beneke:2003zv} is widely
used in literature. 
The BN scheme is constructed for calculations
of twist-3 two parton contributions from a final state pseudo-scalar
meson. 
The BN scheme can not avoid end-point divergences in the hard
spectator and annihilation contributions. 
There exists an ambiguity
that the momentum and coordinate representations for the amplitudes
are used in the calculations. 
Besides, the BN scheme is also used
in the $k_{T}$ factorization \cite{Kurimoto:2001zj,Huang:2004hw}. 
However, the distinguishing differences
between the collinear factorization and the $k_{T}$ factorization
implies that these calculations require some cares. 
The reason is
that the transverse parton momenta are assumed of order $O(\Lambda_{\text{QCD}})$
in the collinear factorization while they are not limited in the $k_{T}$
factorization. 
More detailed comparisons between the BN scheme and the collinear expansion refer to \cite{Yeh:2007fi}.

\subsection{Comparisons with LQCD}
The lattice QCD approach can only calculate the form factors at large
$q^{2}$ due to limits on the inverse space length of the $\pi$
meson energy. 
On the other hand, PQCD approach is applicable for small
$q^{2}$, where the virtual radiations are perturbative. 
Due to $\xi^R$,
the $C^f$ approach developed in this paper for the form factors has extended the application range
from low $q^{2}$ to moderate $q^{2}$. 
These two approaches are complementary
and can be combined to derive form factors of full range $q^{2}$. 
The related works are left in our future publications.
The comparisons between the extrapolation of $F_+^{C^f}(q^2)$ to large $q^2$ region with the 
LQCD calculations by FNAL \cite{Bowler:1999xn, Okamoto:2004xg, Shigemitsu:2004ft, Bailey:2008wp}
and HPQCD \cite{Dalgic:2006dt} collaborations 
show that the $C^f$ prediction is close 
to the FNAL calculations (with a $\chi^2/N=54.8/15$) 
but has a large deviation with the HPQCD calculations (with a $\chi^2/N=271.1/7$).
The $C^f$ prediction is close to the FNAL calculations than the HPQCD calculations.

\section{Discussion and Conclusions}
In this paper, we have proposed a $\xi$-regularization for the logarithmic
end-point divergences in $B\to\pi$ form factors. It can effectively
resolve the end-point divergent problem. The linear end-point divergences
are solved by an simultaneous use of the $\xi$-regularization and
non-constant twist-3 distribution amplitudes. 
The $\xi$-regularization and the existence of a non-constant pseudos-scalar DA are followed
by the collinear expansion. 
The factorizability of $B\to\pi$ form factors is shown up-to $O(\alpha_{s}/m_{B})$ by including complete
twist-3 contributions from the $\pi$ meson and partial $O(1/m_{B}^{2})$ from the $B$ meson. 
The calculations are given explicitly. 
The form factors are calculated and applied to determine the CKM matrix element
$|V_{ub}|$ according to the world average branching fraction.
Eq. (\ref{eq:result-full}) is our main result.   
The result is also applied to make a prediction for the $q^2$ spectrum of $\bar{B}\to\pi l\bar{\nu}_l$.
The predicted differential rate is inconsistent with the BaBar spectrum data.
This discrepancy deserves further studies.
Generalization of our result to other form factors are straightforward.
We leave these interesting tasks to other places.

The twist-3 contributions calculated in the $k_T^f$ and LCSR approaches have been compared to our calculations.
We argue that the twist-3 contributions in these two methods could be overestimated.
If the twist-3 contributions in these two methods can be reduced by an appropriate power expansion method,
it is possible to have a consistent result with our calculations.
However, before the method has been derived under the $k_T^f$ or LCSR formalisms,
the $k^f_T$ and LCSR form factors are larger than the $C^f$ form factor. 

One important application of the present work is to improve the QCD factorization at leading order in $1/m_b$ expansion.
The QCD factorization at leading order in $1/m_b$ expansion demonstrates that the matrix element 
$\langle P_{1}P_{2}|Q_{i}|B\rangle$
for $B\to PP$ decays can be expressed by the factorization formula
\begin{eqnarray}
\langle P_{1}P_{2}|Q_{i}|B\rangle 
& = & 
F^{BP_{1}}
 \int d\xi
  \int du_{1}
   \text{Tr}[T^{I}_{i}(\xi,u_{1})\phi_{B}(\xi)\phi_{P_{2}}(u_{1})]
+(P_{1}\leftrightarrow P_{2})\\
&  & 
+\int d\xi
  \int du_{1}
   \int du_{2}
    \text{Tr}[T^{II}_{i}(\xi,u_{1},u_{2})\phi_{B}(\xi)\phi_{P_{2}}(u_{1})\phi_{P_{2}}(u_{2})]
\,,\nonumber 
\end{eqnarray}
where the $T^{I,II}$ are the type-I and type-II hard scattering kernels
for non-factorizable radiative corrections. 
Due to the end-point divergences
in the form factor $F^{BP_{1}}$ and in twist-3 contributions, the
factorization formula is only valid at leading twist (twist-2). 
The form factor $F^{BP_{1}}$ is isolated from the factorization formula and
identified to be a physical form factor, which can only be determined
by experiments. 
According to the results derived in this paper and
those obtained in \cite{Yeh:2007fi}, 
we propose a generalized QCD factorization
formula, which is valid at twist-3 order. 
The generalized QCD factorization
formula for $B\to PP$ decays with $PP$ two light pseudo-scalar mesons
is given (under two parton approximation) 
\begin{eqnarray}
\langle P_{1}P_{2}|Q_{i}|B\rangle 
& = & f_{P_2}
 \int d\xi
  \int du_{1}
    \text{Tr}[T^{0}_{i}(\xi,u_{1})\phi_{B}(\xi)\phi_{P_{1}}(u_{1})]
    +(P_{1}\leftrightarrow P_{2})\\
&  & 
+ \int d\xi
    \int du_{1}
     \int du_{2}
     \text{Tr}[T^{I}_{i}(\xi,u_{1},u_{2})\phi_{B}(\xi)\phi_{P_{1}}(u_{1})\phi_{P_{2}}(u_{2})]
\nonumber \\
&  & 
+ \int d\xi
   \int du_{1}
    \int du_{2}
     \text{Tr}[T^{II}_{i}(\xi,u_{1},u_{2})\phi_{B}(\xi)\phi_{P_{2}}(u_{1})\phi_{P_{2}}(u_{2})]
\,,\nonumber 
\end{eqnarray}
where $T^{0}$ is the hard scattering kernel for factorizable radiative
corrections, $T^{I}$ is the hard scattering kernel for type-I non-factorizable
radiative corrections, and $T^{II}$ is hard scattering kernel for
type-II non-factorizable radiative corrections. 
The $\xi$-regularization
is used to regularize the end-point divergences in form factors, the
twist-3 annihilation contributions, and twist-3 hard spectator contributions
in charmless $B\to PP$ decays \cite{Yeh:2007fi}. 
The leading order is $O(\alpha_{s})$. 
The improvement in this generalized QCD factorization
is that the form factors are calculable and the precision order is
generalized to include $O(1/m_{B})$ corrections under two parton
approximation. 
The collinear expansion is a necessary tool to derive
correct twist-3 contributions in the above factorization formula.
The application and the formal proof of the generalized QCD factorization
formula will be given elsewhere. 

The generalized QCD factorization formula is valid at twist-3 under
two parton approximation. 
The factorization formula of the same twist
order have also been shown to exist in the $k_{T}$ factorization
and in the SCET, respectively. 
A consistent picture from three different approaches is obtained that the factorization theorem for $B\to PP$
decays is valid up-to twist-3 order under two parton approximation.

\appendix

\section{Reduced hard scattering functions\label{sec:Reduced-hard-scattering}}

The reduced hard scattering function $H_{\mu}^{(1)}(\xi,u)$ is derived
according to the rules given in Section II. According to the Feynman
diagrams depicted in Fig. 1 (a) and (b), the reduced hard scattering
function $H_{\mu}^{(1)}(\xi,u)$ is $H_{\mu}^{(1)}(\xi,u)=H_{\mu}^{(1),(a)}(\xi,u)+H_{\mu}^{(1),(b)}(\xi,u)$
where

\begin{eqnarray}
\left[H_{\mu}^{(1),(a)}(\xi,u)\right]_{ij;kl} 
& = & 
  \frac{C_{F}}{N_{c}}
   \frac{\pi\alpha_{s}}{\eta^{2}E^{4}\xi\bar{u}}
   \left[
         \gamma_{\alpha}\not\! n\not\!\bar{n}
    \right]_{ij}
    \left[
         \gamma_{\beta}
    \right]_{kl}
    d_{\perp}^{\alpha\beta}
    \left(
          P_{B,\mu}-\frac{1}{\eta}p_{\pi,\mu}
    \right)
\,,
\\
\left[
      H_{\mu}^{(1),(b)}(\xi,u)
\right]_{ij;kl} 
& = & 
- \frac{C_{F}}{N_{c}}
  \frac{\pi\alpha_{s}}{\eta^{2}E^{4}\xi\bar{u}(\xi-\eta\bar{u})}
  \left(
        2\bar{\xi}
        \left[
              \not\! n
         \right]_{ij}
         \left[
               \not\! n
         \right]_{kl}
         -\eta\bar{u}
         \left[
               \not\! n\not\!\bar{n}\;\gamma_{\alpha}
         \right]_{ij}
         \left[
               \gamma_{\beta}
          \right]_{kl}
          d_{\perp}^{\alpha\beta}
   \right)
   p_{\pi,\mu}
\,.\nonumber\\
\end{eqnarray}
The reduced hard scattering function 
$H_{\mu\eta\nu}^{(1)}(\xi\,,u_{q}\,,u_{\bar{q}}\,,u_{g})$
are derived according to the Feynman diagrams depicted in Fig.~\ref{fig:fig3}(1)
- (4). The contributions from the other diagrams depicted in 
Fig.~\ref{fig:fig3}(5) - (36) are of higher twist order than twist-3. 
They are neglected in this work. 
The expression for 
$H_{\mu\eta\nu}^{(1)}(\xi\,,u_{q}\,,u_{\bar{q}}\,,u_{g})$
is also written as $H_{\mu\eta\nu}^{(1)}(\xi\,,u_{q}\,,u_{\bar{q}}\,,u_{g})
=H_{\mu\eta\nu}^{(1),(a)}(\xi\,,u_{q}\,,u_{\bar{q}}\,,u_{g})
+H_{\mu\eta\nu}^{(1),(b)}(\xi\,,u_{q}\,,u_{\bar{q}}\,,u_{g})$
where
\begin{eqnarray}
\left[
      H_{\mu\eta\nu}^{(1),(a)}(\xi\,,u_{q}\,,u_{\bar{q}}\,,u_{g})
\right]_{ij;kl} 
& = & 
 \frac{C_{F}}{N_{c}}
  \frac{\pi\alpha_{s}}
   {p_{\pi}^{2}\eta^{2}E^{4}u_{q}u_{\bar{q}}u_{g}\xi\bar{u}_{g}}
   \left[
        \gamma_{\nu}\gamma_{\eta}\gamma_{\alpha}\not\! n\not\!\bar{n}
    \right]_{ij}
    \left[
         \gamma_{\beta}
     \right]_{kl}
     d_{\perp}^{\alpha\beta}
     \left(
           P_{B,\mu}-\frac{1}{\eta}p_{\pi,\mu}
     \right)
\,,
\nonumber\\\\
\left[
     H_{\mu\eta\nu}^{(1),(b)}(\xi\,,u_{q}\,,u_{\bar{q}}\,,u_{g})
\right]_{ij;kl} 
& = & 
  -\frac{C_{F}}{N_{c}}
    \frac{\pi\alpha_{s}}{p_{\pi}^{2}\eta^{2}E^{4}u_{q}u_{\bar{q}}u_{g}\xi(\xi-\eta u_{\bar{q}})}
\nonumber\\
&  & 
  \times
   \left(
         2\bar{\xi}
         \left[
                \gamma_{\nu}\gamma_{\eta}\not\! n
          \right]_{ij}
          \left[
                \not\! n
          \right]_{kl}
          - \eta u_{\bar{q}}
          \left[
                \gamma_{\nu}\gamma_{\eta}\not\! n\not\!\bar{n}\;\gamma_{\alpha}
           \right]
           \left[
                \gamma_{\beta}
           \right]_{kl}
           d_{\perp}^{\alpha\beta}
    \right)
     p_{\pi,\mu}
\,.
\end{eqnarray}

\section{$B$ meson distribution amplitudes}

When subleading order effects in $1/m_{B}$ expansion are considered,
the $B$ meson DAs $\phi_{B}^{tw2}(\xi)$ and $\bar{\phi}_{B}(\xi)$
are defined as \cite{Grozin:1996pq}
\begin{eqnarray}
\langle0|\bar{q}_{j}(\frac{\lambda}{E}\bar{n})b_{i}(0)|\bar{B}\rangle 
& = & 
 \frac{if_{B}}{4N_{c}}
  \left(
     \left(
          \not\! P_{B}+m_{B}
     \right)
     \int_{0}^{1}d\xi e^{i\lambda\xi}
     \left[
           \phi_{B}^{tw2(I)}(\xi)+E(\not\! n-\not\!\bar{n})\bar{\phi}_{B}(\xi)
     \right]
     \gamma_{5}
   \right)_{ij}
\\
&&+\cdots
\,,\nonumber
\end{eqnarray}
where 
$\phi_{B}^{tw2(I)}(\xi)$ 
and 
$\bar{\phi}_{B}(\xi)$ 
correspond
to the $\phi_{B}^{+}(\xi)$ and $\phi_{B}^{+}(\xi)-\phi_{B}^{-}(\xi)$
defined in the literature. 
The models for $\phi_{B}^{+}(\xi)$ and
$\phi_{B}^{-}(\xi)$ are assumed to be \cite{Grozin:1996pq}
\begin{eqnarray}
\phi_{B}^{+}(\xi) 
& = & 
  \frac{m_{B}^{2}}{\omega_{B}^{2}}\xi\exp(-\frac{m_{B}}{\omega_{B}}\xi)
\,,
\\
\phi_{B}^{-}(\xi) 
& = & 
  \frac{m_{B}}{\omega_{B}}\exp(-\frac{m_{B}}{\omega_{B}}\xi)
\,.
\end{eqnarray}
Their first two inverse moments should satisfy the conditions
\begin{eqnarray}
\int_{0}^{1}d\xi\phi_{B}^{(+,-)}(\xi) & = & 1
\,,
\\
\int_{0}^{1}d\xi\frac{\phi_{B}^{(+,-)}(\xi)}{\xi} 
& = & 
  \frac{m_{B}}{\lambda_{B}}\leq\frac{3}{2}\frac{m_{B}}{\bar{\Lambda}}
\,,
\end{eqnarray}
where 
$\bar{\Lambda}=m_{B}-m_{b}$. 
In this work, we employ the
following models for 
$\phi_{B}^{tw2(I)}$ and 
$\bar{\phi}_{B}$
\begin{eqnarray}
\phi_{B}^{tw2(I)}(\xi) 
& = & 
  \phi_{B}^{+}(\xi)=\frac{m_{B}^{2}}{\omega_{B}^{2}}\xi\exp(-\frac{m_{B}}{\omega_{B}}\xi)
\,,
\\
\bar{\phi}_{B}(\xi) 
& = & 
  \phi_{B}^{+}(\xi)-\phi_{B}^{-}(\xi)
 =\frac{m_{B}^{2}}{\omega_{B}^{2}}
   \left(
          \xi-\frac{\omega_{B}}{mB}
   \right)
    \exp(-\frac{m_{B}}{\omega_{B}}\xi)
    \theta(\xi-\frac{\omega_{B}}{mB})
\,,
\end{eqnarray}
where $\theta$ is a step function to insure that $\bar{\phi}_{B}(\xi)$
is positive. 

\bibliographystyle{revtex4/apsrev}
\bibliography{BpiFF-xi-regularization}

\end{document}